\documentclass[iop,revtext4]{emulateapj}
\usepackage{epsfig}
\usepackage{amsmath}

\slugcomment{Accepted for  Publication in The Astronomical Journal}

\shorttitle{NGC 7789: An Open Cluster Case Study}
\shortauthors{Overbeek et al.}

\begin{document}

\title{NGC 7789: An Open Cluster Case Study}
\author{Jamie C. Overbeek}
\affil{Indiana University Astronomy Department, Swain West 319, 727 East 3rd Street, Bloomington, IN 47405, USA; joverbee@indiana.edu}

\author{Eileen D. Friel}
\affil{Indiana University Astronomy Department, Swain West 319, 727 East 3rd Street, Bloomington, IN 47405, USA}

\author{Heather R. Jacobson}
\affil{Massachusetts Institute of Technology, Kavli Institute for Astrophysics and Space Research, Cambridge, MA 02139, USA}

\author{Christian I. Johnson}
\affil{Harvard-Smithsonian Center for Astrophysics, Cambridge, MA 02138, USA}

\author{Catherine A. Pilachowski and Szabolcs M\'{e}sz\'{a}ros}
\affil{Indiana University Astronomy Department, Swain West 319, 727 East 3rd Street, Bloomington, IN 47405, USA}

\begin{abstract}

We have obtained high-resolution spectra of 32 giants in the open cluster NGC 7789 using the Wisconsin--Indiana--Yale--NOAO Hydra spectrograph. We explore differences in atmospheric parameters and elemental abundances caused by the use of the linelist developed for the Gaia-ESO Survey (GES) compared to one based on Arcturus used in our previous work. [Fe/H] values decrease when using the GES linelist instead of the Arcturus-based linelist; these differences are probably driven by systematically lower ($\sim$--0.1 dex) GES surface gravities. Using the GES linelist we determine abundances for 10 elements---Fe, Mg, Si, Ca, Ti, Na, Ni, Zr, Ba, and La. We find the cluster's average metallicity [Fe/H] = 0.03 $\pm$ 0.07 dex, in good agreement with literature values, and a lower [Mg/Fe] abundance than has been reported before for this cluster (0.11 $\pm$ 0.05 dex). We also find the neutron-capture element barium to be highly enhanced---[Ba/Fe] = +0.48 $\pm$ 0.08---and disparate from cluster measurements of neutron-capture elements La and Zr (-0.08 $\pm$ 0.05 and 0.08 $\pm$ 0.08, respectively). This is in accordance with recent discoveries of supersolar Ba enhancement in young clusters along with more modest enhancement of other neutron-capture elements formed in similar environments.

\end{abstract}

\keywords{Galaxy: abundances, open clusters and associations: individual (NGC7789), stars: abundances}

\section{Introduction}

\indent Open clusters (OCs) are important testing grounds for models of stellar evolution and galactic nucleosynthesis. They probe the stellar disk at a wide range of ages and Galactocentric distances (e.g., Janes 1979; Yong et al. 2005; Carraro et al. 2007; Jacobson et al. 2011) and, because their populations have a single age, distance, and metallicity, stellar physical parameters (T$_{\mathrm{eff}}$, log g, etc.) are more easily determined than for field stars. Also, their chemical compositions can be determined with relatively high accuracy through analysis of a large sample of member stars. 
\\
\indent OCs are ideal objects in which to study mixing in stellar interiors; any differences in abundance of stars along the red giant branch (RGB) compared to unevolved dwarfs may be due to such effects, but complicated processes such as internal gravity waves, magnetic-buoyancy-induced mixing, and rotational mixing must be taken into account (Montalb\'{a}n $\&$ Schatzman 2000; Young et al. 2003; Palmerini et al. 2011, etc.). However, a uniform set of abundance measurements for a large number of OCs is not yet available; the heavy elements in particular are relatively understudied. 
\\
\indent The Gaia-ESO Survey (GES) is addressing this gap by providing high-resolution spectra from VLT FLAMES of $\sim$100,000 stars in all components of the Galaxy (Gilmore et al. 2012). Abundances will be determined in a uniform fashion for at least 13 elements (Na, Mg, Si, Ca, Ti, V, Cr, Mn, Fe, Co, Sr, Zr, and Ba) in all cluster and field stars with the possibility of many additional elements---particularly from the higher resolution UVES spectra. Radial velocities from these spectra will complement Gaia astrometry and photometry releases, and the survey's large and homogenous data set will establish a useful new standard in the study of Galactic chemical enrichment. 
\\
\indent All abundances are measured independently by a number of groups whose results are then merged to a uniform system (see Smiljanic et al. 2014). Each group within the collaboration uses their own methods; for example, the EPINARBO (ESO, Padova, Indiana, Arcetri, and Bologna) group uses a pipeline in which equivalent widths are measured with an automated version of DAOSPEC, then input to another program called FAMA that iterates photometric parameters in MOOG automatically (see Magrini et al. 2013; Cantat-Gaudin et al. 2014). Survey collaborators have also developed a single linelist that is used for all analyses (Heiter et al. 2014, in preparation). The GES linelist is more comprehensive than linelists typically seen in individual studies in the literature, and it contains laboratory-determined log(gf) values instead of inverse-solar measurements. 
\\
\indent Jofr\'{e} et al. (2014) have calculated [Fe/H] values for a set of benchmark stars using equivalent width measurements of lines from the new GES linelist, and directly calculating atmospheric parameters from angular diameters, bolometric fluxes, and masses of the benchmark stars. They have generally found good agreement between literature and GES linelist-based [Fe/H] values, with 28 out of 33 benchmark stars yielding results consistent with the literature. 
\\
\indent Sousa et al. (2014) have examined the effects of the GES linelist on spectroscopically determined parameters and Fe abundance, and found that the use of the GES linelist as compared to other linelists resulted in minimal spectroscopic temperature changes ($<$100 K), generally small changes in gravities ($<$ 0.2 dex except for a couple of outliers where $\Delta$log(\textit{g}) $\sim$ 0.5 dex), and changes in [Fe/H] $<$ 0.1 dex. However, there is still more work to do in understanding the effects of the GES pipeline(s) on Fe abundances; also, it is not yet clear how non-Fe abundances measured using this new linelist may differ from values in the literature. One goal of this work is to begin the process of bringing OC literature results onto the GES abundance scale, and to determine possible systematic effects between GES and other large data sets.
\\
\indent The best OCs for a comparison between GES and literature abundance scales would be those that have multiple existing abundance measurements and are populous enough to allow for a large sample of member stars. NGC 7789 (Mel 245, C2354+564) is one such cluster, with a well-defined red clump and RGB and a handful of abundance studies in the literature. At 9.6 kpc from the Galactic center (Jacobson et al. 2011, hereafter JPF11) and roughly solar metallicity (see JPF11, Pancino et al. 2010, hereafter PCRG10, Tautvai\u{s}ien\.{e} et al. 2005, hereafter T05), NGC 7789 has properties typical of an OC. Also, its intermediate age (for an OC) of $\sim$1.6 Gyr old (Gim et al. 1998b) covers a gap in open cluster measurements of neutron-capture trends with age and R$_{\mathrm{GC}}$ (Jacobson \& Friel 2013). Neutron-capture elements are formed by slow (s-process) or rapid (r-process) addition of neutrons onto Fe-peak seed nuclei; they are produced by stars in a large range of masses, and their abundance patterns provide information about the Galactic mass function (Busso et al. 1999; Sneden et al. 2008). 
\\
\indent In this paper, we present abundance measurements of 32 cluster giants for a set of $\alpha$, Fe-peak, and neutron-capture elements. Our spectra cover a wavelength range that includes 40 Fe lines, several Mg, Si, Ca, and Ti lines, and the Na $\lambda$6154/6160 doublet. We are also able to measure one feature each of neutron-capture elements Ba and La, and several Zr lines. This allows us to measure eight elements included in the GES, and form a good basis for comparison of OC results.
\\

\section{Observational Data}
\subsection{Target Selection}

\indent Our target objects were selected to have high membership probabilities ( $>75\%$), based on a previous proper motion study by McNamara \& Solomon (1981).  We also selected for magnitudes/colors consistent with the red clump and RGB using V and I photometry from Gim et al. (1998b). NGC 7789's mass of $\sim$7000$M_{\odot}$ (Wu et al. 2009) and well defined CMD allowed us to select stars along the giant branch that vary over 3 mag in the V band. We also included a few stars that fall in the correct location on a CMD but are suspected to be field stars due to inconsistent proper motions; because of the instrument configuration, some fibers would have been left unused otherwise. We observed 39 stars in total. Figure 1 shows the CMD for NGC 7789 with all $\sim$16,000 cluster stars in Gim et al. (1998b) marked in black, and the stars we observed marked in red (members) and blue (non-members). Membership was determined based on radial velocities, proper motions, and binary status from previous studies (see Section 3).

\begin{figure}
\epsscale{1.1}
\plotone{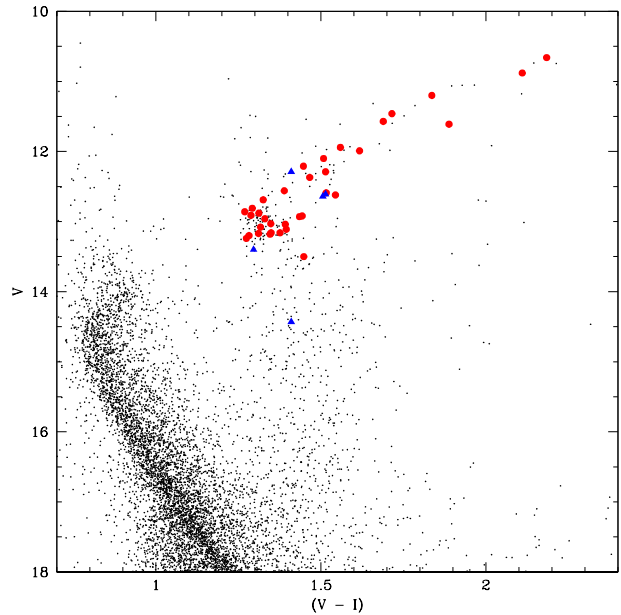}
\caption{Color--magnitude diagram of all N7789 stars (black dots), our sample cluster members (red circles), and sample non-members (blue triangles). Photometry from Gim et al. (1998b).}
\label{CMD}
\end{figure}

\subsection{Observations}

\indent Our data were taken using the Hydra multi-fiber positioner and Bench Spectrograph on the Wisconsin--Indiana--Yale--NOAO (WIYN)\footnote{The WIYN Observatory is a joint facility of the University of Wisconsin-Madison, Indiana University, Yale University, and the National Optical Astronomy Observatory.} 3.5m, on the nights of 2011 August 20--22. A total of two hours of data were taken in four 1800s exposures (one exposure on each of the first two nights and two exposures on the final night); all objects were observed simultaneously. Data were recorded with the STA-1 CCD (2600 x 4000 12 $\mu$m pixels). The red fiber cable was used; the echelle grating in 9th order along with the X18 filter resulted in a wavelength range of $\sim$6050--6380 $\mathrm{\AA}$ with an effective resolution of R = $\lambda$/$\Delta$$\lambda$ $\sim$20,000. This wavelength range was chosen due to the number of features of interest it contains. Our data covers a large number of Fe lines for fundamental analysis and determination of atmospheric parameters, and several lines of $\alpha$ elements Mg, Si, Ca and Ti which provide information about massive star formation rates in the Galaxy. The Na $\lambda$6154/6160 doublet is also in this range; Na abundances have implications for stellar modeling and mixing parameters (Smiljanic 2012). Neutron-capture elements are also included: Ba and La features at $\lambda$6141 and $\lambda$6262 respectively, and several Zr lines. This wavelength range also covers an [O $\textrm{I}$] feature at $\lambda$6300 which we were unfortunately unable to analyze due to interference from night sky emission.
\\
\indent Table 1 lists information for all stellar targets. Star numbering in the first column is from Gim et al. (1998b) and WEBDA\footnote{see http://www.univie.ac.at/webda}; this numbering scheme will be used for the rest of the paper. The second column contains the K\"{u}stner number for each star where available (K\"{u}stner 1923). The table also contains the 1950B coordinates, colors from Gim et al. (1998b) (V, V--I) and 2MASS\footnote{http://www.ipac.caltech.edu/2mass/} (V--K, J--H, J--K), the signal-to-noise ratio (S/N), radial velocity, and membership status. S/N values are based on measurements with SPLOT in IRAF\footnote{IRAF is distributed by the National Optical Astronomy Observatory, which is operated by the Association of Universities for Research in Astronomy, Inc., under cooperative agreement with the National Science Foundation.} in the region of $\lambda$6066-6076; the median S/N is $\sim$100. Radial velocities and membership determination are discussed in Section 3. Also note that photometry in Table 1 is not extinction corrected.

\begin{deluxetable*}{l c c c c c c c c c c c}
\tabletypesize{\small}
\tablewidth{0pt}
\tablecolumns{12}
\tablecaption{Properties of Sample Stars}
\tablehead{\colhead{ID\tablenotemark{a}} & \colhead{ID\tablenotemark{b}} & \colhead{$\alpha_{\mathrm{1950B}}$} & \colhead{$\delta_{\mathrm{1950B}}$} & \colhead{V\tablenotemark{c}} & \colhead{V--I\tablenotemark{c}} & \colhead{V--K\tablenotemark{d}} & \colhead{J--H\tablenotemark{d}} & \colhead{J--K\tablenotemark{d}} & \colhead{S/N\tablenotemark{e}} & \colhead{V$_r$ (km s$^{-1}$)} & \colhead{Member?}}
\startdata
2075 & 14 & 23 53 36.35 & +56 19 23.8 & 13.169 & 1.312 & 2.863 & 0.541 & 0.651 & 81 & -53.83 & M \\
2427 & 47 & 23 53 42.56 & +56 31 59.2 & 13.036 & 1.393 & 3.131 & 0.590 & 0.755 & 51 & -55.39 & M \\
3569 & 136 & 23 53 57.67 & +56 29 01.8 & 13.401 & 1.296 & \nodata & \nodata & \nodata & 72 & -62.20 & NM \\
3798 & 152 & 23 54 00.41 & +56 16 16.8 & 11.941 & 1.559 & 3.480 & 0.693 & 0.859 & 114 & -53.73 & M \\
3835 & 160 & 23 54 01.44 & +56 27 51.1 & 13.034 & 1.348 & 2.984 & 0.602 & 0.695 & 76 & -66.17 & Bin. M \\
4216 & 193 & 23 54 06.34 & +56 32 58.5 & 12.606 & 1.514 & \nodata & \nodata & \nodata & 46 & -60.31 & NM \\
4593 & 232 & 23 54 11.38 & +56 34 05.5 & 13.107 & 1.395 & 3.092 & 0.584 & 0.718 & 87 & -53.25 & M \\
4751 & 244 & 23 54 12.83 & +56 26 11.5 & 13.160 & 1.349 & 3.005 & 0.548 & 0.697 & 55 & -53.01 & M \\
5237 & 297 & 23 54 18.80 & +56 32 38.8 & 12.811 & 1.293 & 2.925 & 0.535 & 0.674 & 104 & -63.05 & Bin. M \\
5594 & 319 & 23 54 22.90 & +56 24 37.9 & 12.864 & 1.269 & 2.823 & 0.531 & 0.685 & 116 & -45.93 & Bin. M \\
5775 & 337 & 23 54 24.80 & +56 21 27.3 & 12.289 & 1.410 & \nodata & \nodata & \nodata & 109 & -34.91 & NM \\
5837 & 353 & 23 54 25.74 & +56 28 45.0 & 12.592 & 1.517 & 3.429 & 0.682 & 0.855 & 105 & -54.65 & M \\
6345 & 415 & 23 54 31.43 & +56 29 15.9 & 10.665 & 2.184 & 4.775 & 0.858 & 1.142 & 210 & -53.91 & M \\
6810 & 468 & 23 54 36.19 & +56 26 39.5 & 11.612 & 1.889 & 4.346 & 0.770 & 0.961 & 197 & -55.02 & M \\
6863 & 476 & 23 54 37.10 & +56 32 08.3 & 13.177 & 1.346 & 2.976 & 0.566 & 0.711 & 99 & -54.66 & M \\
7091 & 501 & 23 54 39.25 & +56 27 47.1 & 11.201 & 1.836 & 4.144 & 0.814 & 1.047 & 188 & -54.69 & M \\
7369 & 549 & 23 54 42.11 & +56 24 16.5 & 12.290 & 1.514 & 3.440 & 0.679 & 0.856 & 150 & -53.77 & M \\
7617 & 575 & 23 54 45.00 & +56 29 37.8 & 12.100 & 1.509 & 3.434 & 0.665 & 0.846 & 163 & -54.70 & M \\
7640 & 580 & 23 54 45.36 & +56 28 51.2 & 13.203 & 1.282 & 2.927 & 0.550 & 0.683 & 92 & -55.22 & M \\
7691 & 579 & 23 54 45.36 & +56 20 06.1 & 14.432 & 1.410 & \nodata & \nodata & \nodata & 52 & -88.83 & NM \\
7867 & 601 & 23 54 47.17 & +56 19 14.1 & 12.373 & 1.467 & 3.299 & 0.644 & 0.809 & 66 & -55.33 & M \\
8061 & 626 & 23 54 49.13 & +56 18 20.6 & 12.962 & 1.331 & 2.906 & 0.576 & 0.695 & 70 & -54.95 & M \\
8293 & 669 & 23 54 52.51 & +56 31 48.3 & 11.456 & 1.716 & 3.915 & 0.744 & 1.032 & 176 & -54.23 & M \\
8316 & 665 & 23 54 52.21 & +56 22 54.1 & 12.913 & 1.289 & 2.853 & 0.519 & 0.677 & 52 & -57.87 & M \\
8734 & 732 & 23 54 57.55 & +56 25 41.4 & 12.694 & 1.325 & 2.974 & 0.565 & 0.725 & 114 & -53.44 & M \\
8799 & 751 & 23 54 58.77 & +56 34 09.3 & 10.878 & 2.110 & 4.625 & 0.939 & 1.167 & 298 & -57.95 & Bin. M \\
9010 & 774 & 23 55 00.60 & +56 27 23.5 & 13.242 & 1.274 & 2.829 & 0.514 & 0.640 & 84 & -54.50 & M \\
9728 & 859 & 23 55 08.48 & +56 28 41.1 & 12.639 & 1.507 & \nodata & \nodata & \nodata & 78 & -35.27 & NM \\
10133 & \nodata & 23 55 13.70 & +56 33 59.2 & 13.164 & 1.377 & 3.031 & 0.542 & 0.705 & 98 & -54.58 & M \\
10578 & \nodata & 23 55 17.89 & +56 26 23.1 & 12.884 & 1.312 & 2.956 & 0.555 & 0.717 & 125 & -55.15 & M \\
10584 & \nodata & 23 55 17.29 & +56 18 08.4 & 13.502 & 1.449 & 3.231 & 0.619 & 0.790 & 67 & -55.79 & M \\
10645 & \nodata & 23 55 18.29 & +56 22 40.7 & 13.084 & 1.317 & 2.942 & 0.567 & 0.690 & 89 & -55.69 & M \\
10740 & \nodata & 23 55 19.98 & +56 30 15.5 & 12.620 & 1.544 & 3.501 & 0.763 & 0.885 & 130 & -53.98 & M \\
10996 & \nodata & 23 55 22.83 & +56 27 50.2 & 12.208 & 1.448 & 3.307 & 0.671 & 0.822 & 144 & -54.92 & M \\
11413 & 1066 & 23 55 27.95 & +56 33 30.4 & 11.986 & 1.618 & 3.653 & 0.773 & 0.955 & 198 & -54.34 & M \\
11573 & \nodata & 23 55 29.08 & +56 22 39.9 & 12.557 & 1.389 & 3.141 & 0.561 & 0.759 & 119 & -55.27 & M \\
11622 & \nodata & 23 55 30.50 & +56 30 38.7 & 12.933 & 1.436 & 3.206 & 0.656 & 0.776 & 96 & -54.31 & M \\
12478 & 1147 & 23 55 42.38 & +56 35 04.4 & 11.576 & 1.689 & 3.797 & 0.832 & 1.043 & 233 & -56.18 & M \\
12550 & \nodata & 23 55 43.25 & +56 32 30.3 & 12.918 & 1.443 & 3.158 & 0.650 & 0.780 & 96 & -52.95 & M \\
\enddata
\tablenotetext{a}{ID numbers from Gim et al. (1998b), WEBDA (used in all further tables).}
\tablenotetext{b}{ID numbers from K\"{u}stner (1923), if available.}
\tablenotetext{c}{V and V-I photometry from Gim et al. (1998b)}
\tablenotetext{d}{J, H, and K photometry from 2MASS}
\tablenotetext{e}{Per pixel, measured from the final averaged spectrum around $\lambda$6072.}
\label{star_info}
\end{deluxetable*}

\subsection{Data Reduction}

\indent Spectra were reduced following standard IRAF procedures. Exposures were first bias subtracted using CCDPROC, then spectra were extracted with the DOHYDRA routine. DOHYDRA does flat-fielding, dispersion correction, and sky subtraction on WIYN-Hydra images. The exposures were then averaged with sigma-rejection to remove cosmic rays, and the combined spectra were continuum-fit. Our data included several nightly telluric standards, which were later used to correct spectra for atmospheric absorption in the IRAF task TELLURIC. This correction was important for the measurement of Mg lines in the $\lambda$6315-6325 region.

\section{Radial Velocities}

\indent HD 212943 was chosen as a radial velocity standard due to its brightness, V = +4.8, and its similarity in spectral type to the target objects - it is a K0III star which places it in the approximate temperature range of our sample (Fricke et al. 1991). Three exposures of HD 212943 were taken on the last night of observations (August 22). Stellar radial velocities were determined using the IRAF package FXCOR, which performs a Fourier cross-correlation on an object and a template spectrum; the peak of the cross-correlation function is fit with a Gaussian to determine the best velocity. The typical error in a cross-correlation was $\sim$0.7 km s$^{-1}$. RVCORRECT was then used to determine the final heliocentric velocity. Velocities were calculated for each of the four exposures, and results from the two exposures taken on night three were averaged to obtain one velocity per night of observations. Night-to-night variations in radial velocity were less than 1 km s$^{-1}$ for all objects except 9728 (which was determined to be a non-member). Velocities from each night of observation were then averaged to obtain a final velocity for each object; the typical error on the final stellar velocity (the standard deviation of nightly velocities) was $\sim$0.4 km s$^{-1}$.
\\
\indent Although we have proper motion data for our entire star sample, radial velocities provide additional information about membership. Figure 2 shows the radial velocity histogram for all target objects. We are able to directly compare radial velocities with measurements of sample stars in the literature; our sample overlaps with JPF11, PCRG10, Jacobson (2009), and Gim (1998a). Figure 3 shows a comparison plot of velocities found in the literature versus those found in this study. Stars that had $>$ 5 km s$^{-1}$ differences between our measurements and literature values were classified as possible binaries; for non-binary members, differences between velocities found by this study and either comparison study were 0.0 $\pm$ 1.2 km s$^{-1}$.

\begin{figure}
\epsscale{1.1}
\plotone{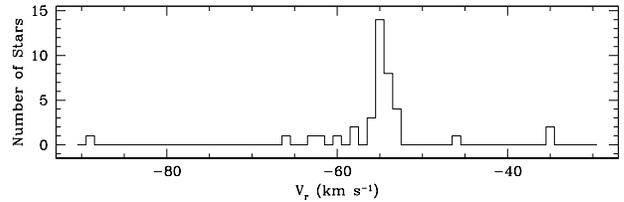}
\caption{\small{Average stellar radial velocity distribution of target objects.}}
\label{RV_hist}
\end{figure}

\indent Membership was determined based on consideration of radial velocities, proper motions, and potential binary status. Proper motion membership probabilities are from McNamara \& Solomon (1981) and radial velocity membership probabilities are based on the average and standard deviation of velocities for all 39 observed stars (--55 $\pm$ 8 km s$^{-1}$) :
\begin{itemize}
\item 7691, at --3$\sigma$ from the cluster average, has a 0.3\% membership probability based on radial velocity. We consider it a non-member.
\item 5775 and 9728 are each about +2.5$\sigma$ from the mean radial velocity with a membership probability of 1\%, and are determined to be non-members.
\item 5594, with a radial velocity of --45.93 km s$^{-1}$, falls outside of the peak of the cluster V$_r$ distribution, but Gim et al. (1998a) find it to have a radial velocity of --57.52 km s$^{-1}$ and McNamara $\&$ Solomon (1981) give it a 98$\%$ probability of being a cluster member, which suggests it is in fact a binary member. 
\item 3835 similarly appears to be outside of the cluster distribution at --66 km s$^{-1}$, but its proper motion is consistent with membership and JPF11 previously determined it to be a binary star, so we include it in our sample as a possible binary member.
\\
\item We have found a discrepant radial velocity for 5237, but McNamara $\&$ Solomon (1981) give it a high membership probability (97$\%$) and PCGR10 find a V$_r$ of --57.17 km s$^{-1}$ consistent with the cluster velocity, so we include it in our sample as a possible binary member. 
\item 4216 and 3569  are more difficult to place at --60 to --62 km s$^{-1}$; considering the distribution of stars with --65 km s$^{-1}$ $<$ V$_r$ $<$ --40 km s$^{-1}$, they are 2.0 and 1.5$\sigma$ from the average giving them membership probabilities based on radial velocities alone of 13\% and 5\%, respectively. 4216 and 3569 also have inconsistent proper motions---McNamara and Solomon (1981) give each a 0\% probability of membership - so we exclude them.
\end{itemize}

\noindent Using non-binary cluster members, we determine the average cluster velocity to be --54.6 $\pm$ 1.0 km s$^{-1}$; the given error is the standard deviation of individual stellar velocities. Our cluster velocity is in excellent agreement with Gim et al.'s (1998a) value of V$_r$ = --54.9 $\pm$ 0.9 km s$^{-1}$ and JPF11's V$_r$ = --54.7 $\pm$ 1.3 km s$^{-1}$. 

\begin{figure}
\epsscale{1.1}
\plotone{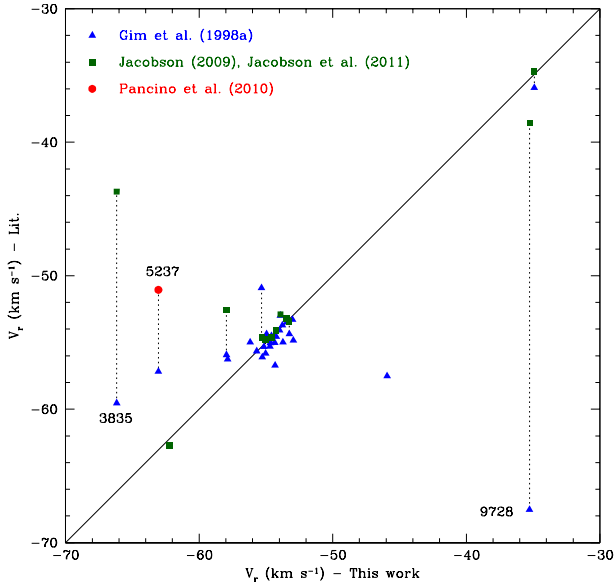}
\caption{\small{Comparison of stellar radial velocities between this study and the literature. Blue triangles are data from Gim et al. (1998a), green squares are from JPF11 and Jacobson (2009), and red circles are from PCRG10. Dotted lines connect literature measurements of the same star, and the solid line indicates a 1:1 correspondence.}}
\label{RV_fig}
\end{figure}

\section{Linelists and Equivalent Widths}

\indent The third internal data release of the GES is now underway, and collaborators are working on moving literature abundance measurements onto the GES scale. One way to gauge the systematic offsets between literature and GES results is to compare measurements made on the same data with different linelists, which we have done for NGC 7789. We adopt the newest version of the GES linelist (Heiter et al. 2014, in preparation) as well as the linelist from Friel et al. (2003), most recently updated in JPF11, which uses log(gf) values inversely determined from a high-resolution spectrum of Arcturus (Hinkle et al. 2000). GES log(gf) values typically compare well with log(gf) values from the Arcturus-based linelist. 
\\
\indent Figure 4 plots $\Delta$log(gf), i.e., GES log(gf) -- Arcturus log(gf), against the Arcturus log(gf) values, with a solid line indicating the best least-squares fit to the data. GES values are on average higher by 0.03 $\pm$ 0.13 dex, which we consider to be an insignificant difference. We 
\\
\\
have conducted separate analyses of stellar equivalent widths for each line list to derive two different sets of atmospheric parameters and Fe abundances (see Section 6). Parameters and abundances derived using lines from the GES linelist will hereafter be labeled simply as the ``GES [param.],'' and those derived using the JPF11 linelist as the ``Arcturus [param.].''

\begin{figure}
\epsscale{1.0}
\plotone{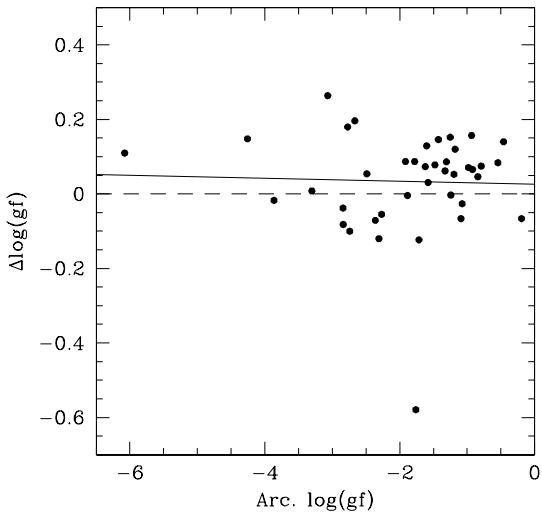}
\caption{\small{Comparison of log(gf) values for all lines used in our study, with GES log(gf) -- Arcturus log(gf) vs. Arcturus log(gf). The dotted line is at zero (no difference) and the solid line marks a least-squared fit to the differences. The point at $\Delta$log(gf) $\sim$ -0.6 is an Fe I line.}}
\label{log_gf}
\end{figure}
\indent Equivalent widths were measured for Si, Ca, Ti, Na, Ni, and Zr by fitting lines with Gaussian profiles. Table 2 lists wavelengths and equivalent widths of all lines used for each star, along with excitation potentials (E.P.) and log(gf) values for both linelists. We have also included in Table 2 Mg equivalent width measurements, which give systematically high abundances, because we use them for comparison purposes with our synthesis measurements (see Sections 6.1, 7.2). All lines were measured multiple times in SPLOT, with an average of $\sim$3 measurements per line staggered over several days. Any lines with equivalent widths $<$ 10 m$\mathrm{\AA}$ or $>$ 170 m$\mathrm{\AA}$ were not used, due to the difficulty of accurately measuring weak lines and the saturation of larger lines. Some lines in each linelist routinely gave equivalent widths $>$ 170 m$\mathrm{\AA}$, or were always badly blended; these lines were not used and do not appear in Table 2. The typical error in equivalent width measurements, based on comparisons between stars with similar atmospheric parameters, was $\sim$5 m$\mathrm{\AA}$.

\begin{deluxetable*}{l c c c c c c}
\tabletypesize{\small}
\tablewidth{0pt}
\tablecolumns{7}
\tablecaption{Linelist Parameters and Equivalent Width Measurements\tablenotemark{a}}
\tablehead{\colhead{Star} & \colhead{$\lambda (\mathrm{\AA})$} & \colhead{El.} & \colhead{E.P.} & \colhead{GES log(gf)} & \colhead{Arc. log(gf)} & \colhead{EQW (m$\mathrm{\AA}$)\tablenotemark{b}}}
\startdata
2075 & 6056.01 & 26.0 & 4.733 & -0.460 & -0.544 & \nodata \\
2075 & 6079.01 & 26.0 & 4.652 & -1.020 & \nodata & 81.6 \\
2075 & 6093.64 & 26.0 & 4.607 & -1.400 & -1.478 & 62.7 \\
2075 & 6094.37 & 26.0 & 4.652 & -1.840 & -1.717 & 57.1 \\
2075 & 6096.66 & 26.0 & 3.984 & -1.830 & -1.917 & 79.5 \\
2075 & 6105.13 & 26.0 & 4.548 & \nodata & -1.994 & 44.9 \\
2075 & 6120.25 & 26.0 & 0.915 & -5.970 & -6.080 & 45.4 \\
2075 & 6127.91 & 26.0 & 4.143 & -1.399 & \nodata & 86.1 \\
2075 & 6151.62 & 26.0 & 2.176 & -3.295 & -3.303 & 103.9 \\
2075 & 6157.73 & 26.0 & 4.076 & -1.160  & -1.094 & 115.7 \\
\enddata
\tablenotetext{a}{Table 2 is published in its entirety in the electronic edition of the AJ. A portion is shown here for guidance regarding its form and content.}
\tablenotetext{b}{Mg equivalent widths are included because they are referenced in comparison to the reported abundances determined via spectral synthesis.}
\label{linelist}
\end{deluxetable*}

\section{Atmospheric Parameters}

\indent The calculation of atmospheric parameters requires information about basic cluster parameters such as distance and reddening. Various photometry studies of NGC 7789 have yielded reddening values E(B--V) from 0.22 (Clari\'{a} 1979) to 0.32 (Strom $\&$ Strom 1970, Mart\'{i}nez-Roger et al. 1994), and distance moduli (m--M)$_{0}$ from 11.0 (Strom $\&$ Strom 1970) to 11.5 (Jennens $\&$ Helfer 1975, Janes 1977). Gim et al. (1998b) remains the largest and most comprehensive photometric study of this cluster, so we have adopted their extinction value E(B--V) = 0.28 and distance modulus (m--M)$_V$ = 12.20. 
\\
\indent We used JHK photometry from 2MASS along with V-band measurements from Gim et al. (1998b) to estimate stellar temperatures following the calibration of Alonso et al. (1999). After calculating the true V--K, J--H, and J--K colors using extinction corrections from Binney \& Tremaine (2008), we used color transformations from Alonso et al. (1998) and Carpenter (2001) and corresponding temperature--color relations from Alonso et al. (1999) to calculate a set of initial temperatures for each star. These individual temperatures were then averaged to yield the final photometric temperature. Gravities were calculated using the standard equation:
\begin{equation}
\begin{split}
\mathrm{log}(g) =  & \mathrm{log}\left(\frac{M}{M_{\odot}}\right) - 0.4(M_{\mathrm{bol}, \odot} - M_{\mathrm{bol}, \ast}) \\ & + 4\mathrm{log}\left(\frac{T}{T_{\odot}}\right) + \mathrm{log}\left(g_{\odot}\right)
\end{split}
\end{equation}

\noindent where log$\left(g_{\odot}\right)$ = 4.44, $T_{\odot}$ = 5770 K,  and $M_{\mathrm{bol}, \odot}$ = 4.72 (Allen 1976). Bolometric corrections for each star were calculated after Alonso et al. (1999), and a turn-off mass of $\sim$2 $M_{\odot}$ was assumed appropriate to the age of the cluster. Finally, an initial microturbulent velocity of 1.5 km s$^{-1}$ was assumed for all stars (JPF11 find a typical microturbulent velocity of 1.5 km s$^{-1}$ for giants).
\\
\indent These photometrically determined atmospheric parameters were then used as a starting point for the spectroscopic determination. Our wavelength range includes 35 Fe I lines and six Fe II lines; the GES linelist includes 35 Fe I and six Fe II lines, while the Arcturus-based linelist includes 21 Fe I and four Fe II lines. This number is sufficient to distinguish abundance trends with line strength and E.P. for Fe I, and also refine log(\textit{g}) values by requiring ionization equilibrium between Fe I and Fe II abundances. Table 3 gives the spectroscopic atmospheric parameters and iron abundances for 32 stars judged to be members. Stars 6345 and 8799 are likely cluster members but are not included in Table 3; these two objects are the coolest in our sample and due to difficulties with continuum fitting over our limited spectral range, the errors on determined iron abundances were $\sim$0.3 dex. Since any subsequent spectroscopic parameter or abundance determinations would be unreliable, we exclude these two stars from the rest of the analysis.

\begin{deluxetable*}{l c c c c c c c c c c c c c c}
\tablewidth{0pt}
\tablecolumns{15}
\tablecaption{Atmospheric Parameters of Member Stars}
\tablehead{\colhead{} & \colhead{GES} & \colhead{GES} & \colhead{GES} & \colhead{GES} & \colhead{GES} & \colhead{GES} & \colhead{GES} & \colhead{Arc.} & \colhead{Arc.} & \colhead{Arc.} & \colhead{Arc.} & \colhead{Arc.} & \colhead{Arc.} & \colhead{Arc.} \\ \colhead{} & \colhead{T} & \colhead{log (g)} & \colhead{vt} & \colhead{[Fe/H]} & \colhead{$\sigma_{\mathrm{Fe}}$} & \colhead{Fe I} & \colhead{Fe II} & \colhead{T} & \colhead{log(g)} & \colhead{vt} & \colhead{[Fe/H]} & \colhead{$\sigma_{\mathrm{Fe}}$} & \colhead{Fe I} & \colhead{Fe II} \\ \colhead{ID\tablenotemark{a}} & \colhead{(K)} & \colhead{(dex)} & \colhead{(km s$^{-1}$)} & \colhead{(dex)} & \colhead{(dex)} & \colhead{no.} & \colhead{no.} & \colhead{(K)} & \colhead{(dex)} & \colhead{(km s$^{-1}$)} & \colhead{(dex)} & \colhead{(dex)} & \colhead{no.} & \colhead{no.}}
\startdata
2075 & 5050 & 2.6 & 1.50 & 0.13 & 0.13 & 31 & 6 & 5050 & 2.7 & 1.45 & 0.22 & 0.18 & 20 & 4 \\
2427 & 4800 & 2.5 & 1.45 & 0.18 & 0.16 & 27 & 6 & 4900 & 2.5 & 1.50 & 0.30 & 0.15 & 20 & 4 \\
3798 & 4450 & 1.8 & 1.55 & -0.06 & 0.16 & 29 & 6 & 4500 & 2.2 & 1.50 & 0.07 & 0.15 & 21 & 4 \\
3835 & 4950 & 2.6 & 1.40 & 0.06 & 0.16 & 35 & 5 & 5000 & 2.4 & 1.50 & 0.09 & 0.16 & 21 & 4 \\
4593 & 4800 & 2.4 & 1.30 & 0.14 & 0.13 & 30 & 6 & 4800 & 2.4 & 1.40 & 0.15 & 0.18 & 19 & 4 \\
4751 & 4950 & 2.3 & 1.60 & 0.07 & 0.15 & 27 & 6 & 4950 & 2.5 & 1.70 & 0.11 & 0.23 & 19 & 4 \\
5237 & 4950 & 2.2 & 1.25 & -0.01 & 0.14 & 32 & 6 & 5050 & 2.4 & 1.35 & 0.06 & 0.18 & 20 & 4 \\
5594 & 5000 & 2.5 & 1.20 & -0.02 & 0.13 & 28 & 6 & 5000 & 2.5 & 1.40 & -0.02 & 0.22 & 19 & 4 \\
5837 & 4550 & 2.0 & 1.45 & 0.02 & 0.12 & 25 & 5 & 4550 & 2.2 & 1.45 & 0.12 & 0.19 & 20 & 4 \\
6810 & 4300 & 1.5 & 1.50 & 0.10 & 0.15 & 18 & 6 & 4175 & 1.4 & 1.60 & 0.06 & 0.14 & 18 & 4 \\
6863 & 4950 & 2.5 & 1.40 & 0.11 & 0.11 & 32 & 6 & 4950 & 2.7 & 1.40 & 0.20 & 0.19 & 20 & 4 \\
7091 & 4200 & 1.3 & 1.50 & -0.02 & 0.18 & 20 & 5 & 4000 & 0.9 & 1.50 & 0.04 & 0.16 & 17 & 4 \\
7369 & 4500 & 1.9 & 1.55 & -0.05 & 0.12 & 25 & 5 & 4550 & 2.2 & 1.55 & 0.07 & 0.13 & 19 & 4 \\
7617 & 4600 & 1.9 & 1.60 & -0.04 & 0.15 & 26 & 5 & 4550 & 2.0 & 1.50 & 0.08 & 0.12 & 18 & 4 \\
7640 & 5050 & 2.6 & 1.55 & 0.10 & 0.15 & 29 & 6 & 5000 & 2.6 & 1.40 & 0.19 & 0.22 & 20 & 4 \\
7867 & 4650 & 2.1 & 1.45 & 0.06 & 0.14 & 29 & 6 & 4700 & 2.5 & 1.50 & 0.23 & 0.13 & 17 & 4 \\
8061 & 4900 & 2.2 & 1.45 & 0.10 & 0.18 & 32 & 6 & 4900 & 2.2 & 1.50 & 0.07 & 0.15 & 18 & 4 \\
8293 & 4200 & 1.4 & 1.55 & -0.01 & 0.19 & 24 & 5 & 4200 & 1.5 & 1.50 & 0.07 & 0.16 & 19 & 4 \\
8316 & 5050 & 2.4 & 1.60 & 0.05 & 0.21 & 28 & 6 & 4850 & 2.0 & 1.40 & 0.05 & 0.23 & 18 & 4 \\
8734 & 4900 & 2.3 & 1.40 & 0.10 & 0.11 & 30 & 6 & 4900 & 2.4 & 1.25 & 0.26 & 0.17 & 20 & 4 \\
9010 & 5050 & 2.7 & 1.50 & 0.10 & 0.11 & 31 & 6 & 5050 & 2.6 & 1.40 & 0.19 & 0.15 & 20 & 4 \\
10133 & 4900 & 2.5 & 1.45 & 0.00 & 0.15 & 33 & 6 & 4900 & 2.8 & 1.35 & 0.14 & 0.21 & 21 & 4 \\
10578 & 4850 & 2.2 & 1.40 & 0.01 & 0.14 & 33 & 6 & 4950 & 2.7 & 1.60 & 0.13 & 0.19 & 21 & 4 \\
10584 & 4750 & 2.5 & 1.30 & 0.10 & 0.11 & 31 & 6 & 4800 & 2.5 & 1.20 & 0.25 & 0.16 & 19 & 4 \\
10645 & 4950 & 2.4 & 1.50 & 0.11 & 0.15 & 29 & 6 & 4950 & 2.7 & 1.45 & 0.26 & 0.18 & 17 & 4 \\
10740 & 4600 & 2.0 & 1.45 & 0.07 & 0.12 & 26 & 6 & 4600 & 2.1 & 1.50 & 0.08 & 0.15 & 19 & 4 \\
10996 & 4650 & 1.8 & 1.50 & -0.09 & 0.10 & 27 & 6 & 4700 & 2.3 & 1.55 & 0.09 & 0.16 & 20 & 4 \\
11413 & 4350 & 1.6 & 1.50 & -0.11 & 0.10 & 23 & 5 & 4325 & 1.6 & 1.60 & -0.09 & 0.12 & 18 & 4 \\
11573 & 4700 & 2.0 & 1.40 & 0.00 & 0.12 & 30 & 5 & 4750 & 2.3 & 1.50 & 0.07 & 0.17 & 19 & 4 \\
11622 & 4750 & 2.3 & 1.45 & 0.04 & 0.18 & 30 & 6 & 4700 & 2.3 & 1.45 & 0.04 & 0.15 & 19 & 4 \\
12478 & 4250 & 1.3 & 1.65 & -0.12 & 0.10 & 20 & 5 & 4250 & 1.5 & 1.55 & 0.08 & 0.13 & 17 & 4 \\
12550 & 4750 & 2.1 & 1.30 & 0.03 & 0.15 & 34 & 6 & 4750 & 2.1 & 1.30 & 0.12 & 0.19 & 21 & 4 \\
\enddata
\tablenotetext{a}{ID numbers from Gim et al. (1998b), WEBDA.}
\label{atm_params}
\end{deluxetable*}

\indent Final spectroscopic parameters were determined using the ABFIND routine in the 2010 version of MOOG (Sneden 1973) along with spherical MARCS model atmospheres (Gustafsson et al. 2008). MOOG 2010 uses solar abundances from Anders $\&$ Grevesse (1989) to calculate all differential abundances. The user then iterates on the initial parameters to remove trends in abundance with excitation potential (for temperature) and equivalent width (for microturbulence), and selects the gravity that satisfies ionization equilibrium for different species of an element. In this way we calculated separate values for all parameters based on each linelist. 
\\
\indent Figure 5 compares spectroscopic determinations of each parameter from the GES and Arcturus-based linelists; each point is a different star, with the dashed line showing a 1:1 correspondence and the solid line showing the least-squares fit to the data. Typical errors are also marked on each plot (see below). Errors for each atmospheric parameter were determined by taking into account (1) the typical differences between parameters found using each line list and (2) the limit for visually identifying trend lines in MOOG. 
\\
\indent Our temperatures display no systematic differences between linelists, with a fairly tight 1:1 correlation between GES and Arcturus-based values. The two linelists give a typical $\Delta$T of $\pm$50K; typical differences between photometric and spectroscopic values were also $\sim$50K and no more than 200K. However, trends in abundance with excitation potential (E. P.) are clearly visible only for temperature differences of 100K or more, so we adopt an error of $\pm$ 100K. We find a slight offset in log(g) values - GES gravities systematically decrease by 0.1 dex from photometric values while Arcturus gravities are roughly photometric, so that GES gravities are $\sim$0.1 dex smaller than found for Arcturus linelists. GES and Arcturus microturbulent velocities are 1:1 within errors, with an rms scatter of 0.1 km s$^{-1}$.
\\
\indent Differences in determined Fe abundances are probably driven by the offset in log(g) values; GES Fe I abundances are $\sim$0.1 dex lower than Arcturus values. Errors for stellar Fe abundances given in Table 3 are the standard deviation of MOOG-calculated abundances for each line of Fe I. GES Fe abundances show a typical error of 0.14 dex per star, and Arcturus Fe abundances have errors of $\sim$0.17 dex. With GES parameters, the cluster [Fe/H] is found to be 0.03 $\pm$ 0.08 dex, whereas Arcturus parameters give [Fe/H] = 0.11 $\pm$ 0.09 dex. Note that [Fe/H] and all other cluster abundances are presented as weighted averages, so that outliers with large errors do not contribute as strongly to the final value; associated uncertainties are the standard deviation of all stellar abundances. 
\\
\indent Our sample overlaps with those of three previous abundance studies of NGC 7789: JPF11, PCRG10, and T05. Although the GES and Arcturus [Fe/H] averages are consistent with each other given their respective errors, the GES metallicity is significantly closer to those found by JPF11 (0.02 dex), T05 (-0.04 dex), and PCRG10 (0.02 dex). In the case of JPF11, atmospheric parameters are not refined spectroscopically, so [Fe/H] values used in the cluster average and subsequent ratio calculations are Fe I abundances. As noted below, our spectroscopically refined parameters using the Arcturus linelist have slightly higher temperatures, which could explain why our Arcturus [Fe/H] value is greater than that of JPF11. Since we have judged our atmospheric parameters derived from the GES linelist to be more trustworthy due to a larger number of Fe lines and smaller errors, we have used this set of parameters for all further abundance analyses. Table 4 lists the uncertainties in elemental abundances due to changes in atmospheric parameters for three stars that span the total parameter space. 

\begin{figure*}
\epsscale{1.0}
\plotone{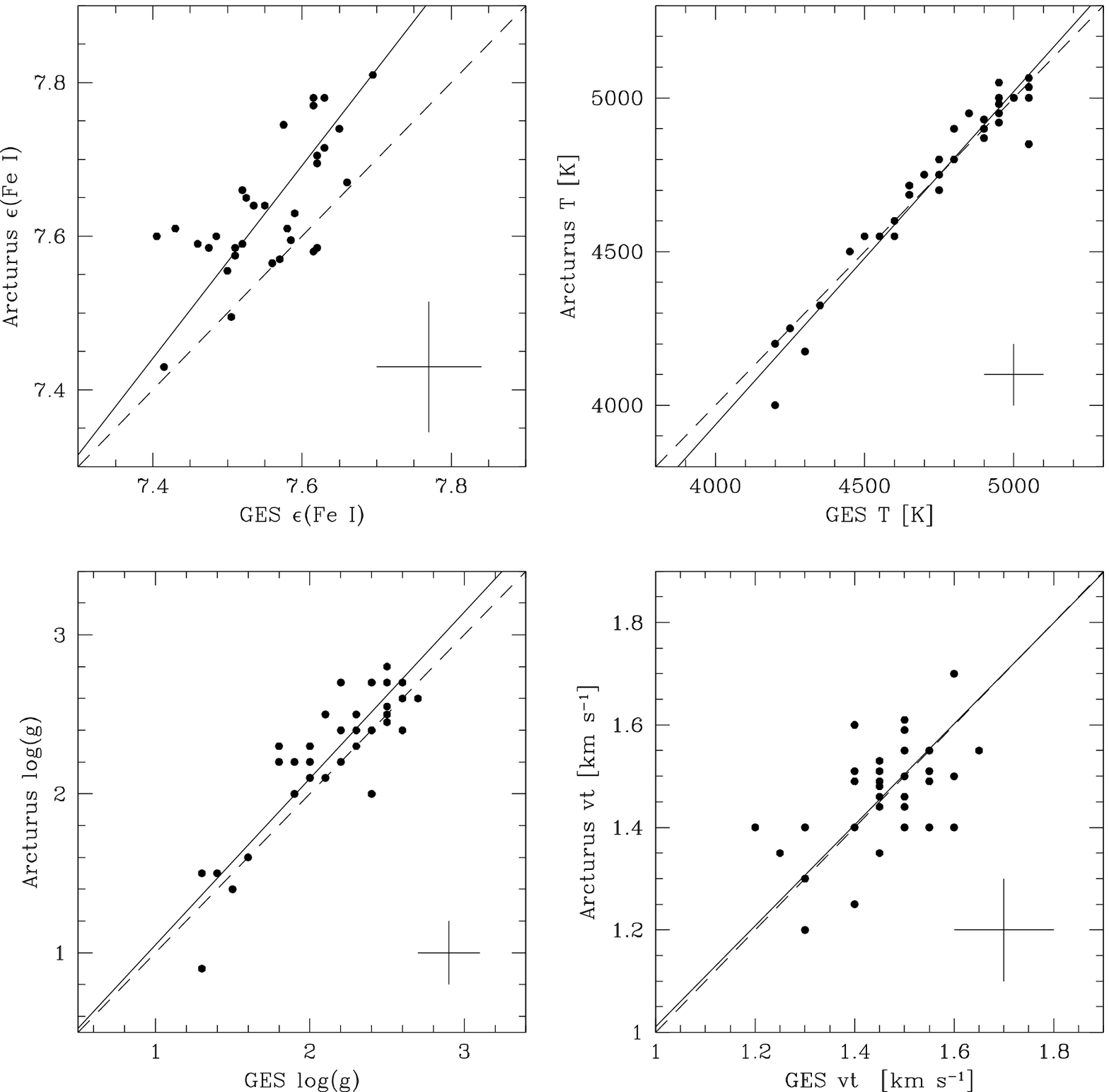}
\caption{\small{Comparison of atmospheric parameters spectroscopically determined using the GES linelist and Arcturus-based linelist. Dashed lines indicate a 1:1 correspondence, and solid lines are a fit to the data. Typical errors are marked on each plot.}}
\label{atm_trends}
\end{figure*}

\begin{deluxetable}{l c c c c c}
\tablewidth{0pt}
\tablecolumns{5}
\tablecaption{Abundance Uncertainties Due to Atmospheric Parameters}
\tablehead{\colhead{Star} & \colhead{[X/H]} & \colhead{T$_{\mathrm{eff}}$ +100K} & \colhead{log(g) +0.2} & \colhead{vt +0.1 km s$^{-1}$}}
\startdata
5594\tablenotemark{a} & Fe I & 0.09 & 0.01 & -0.06 \\
5594 & Mg I & 0.04 & 0.00 & -0.01 \\
5594 & Si I & 0.00 & 0.03 & -0.01 \\
5594 & Ca I & 0.09 & -0.03 & -0.04 \\
5594 & Ti I & 0.13 & 0.00 & -0.03 \\
5594 & Na I & 0.07 & 0.00 & -0.02 \\
5594 & Ni I & 0.07 & 0.02 & -0.05 \\
5594 & Zr I & 0.16 & 0.00 & -0.01 \\
5594 & Ba II & 0.01 & 0.03 & -0.05 \\
5594 & La II & -0.05 & 0.02 & -0.07 \\
7091\tablenotemark{b} & Fe I & 0.00 & 0.05 & -0.07 \\
7091 & Mg I & 0.00 & 0.03 & -0.02 \\
7091 & Si I & -0.09 & 0.06 & -0.02 \\
7091 & Ca I & 0.11 & 0.01 & -0.03 \\
7091 & Ti I & 0.14 & 0.04 & -0.09 \\
7091 & Na I & 0.08 & 0.01 & -0.06 \\
7091 & Ni I & -0.02 & 0.06 & -0.05 \\
7091 & Zr I & 0.18 & 0.04 & -0.09 \\
7091 & Ba II & -0.01 & 0.04 & -0.07 \\
7091 & La II & 0.02 & 0.06 & -0.02 \\
11622\tablenotemark{c} & Fe I & 0.08 & 0.02 & -0.06 \\
11622 & Mg I & 0.04 & 0.01 & -0.01 \\
11622 & Si I & -0.03 & 0.04 & -0.02 \\
11622 & Ca I & 0.10 & -0.03 & -0.05 \\
11622 & Ti I & 0.15 & 0.01 & -0.02 \\
11622 & Na I & 0.09 & 0.00 & -0.03 \\
11622 & Ni I & 0.03 & 0.03 & -0.05 \\
11622 & Zr I & 0.18 & 0.00 & -0.01 \\
11622 & Ba II & 0.05 & 0.08 & -0.03 \\
11622 & La II & -0.04 & 0.08 & -0.01 \\
\enddata
\tablenotetext{a}{Starting parameters for 5594 are T = 5000K, log(g) = 2.5, vt = 1.20 km s$^{-1}$}
\tablenotetext{b}{Starting parameters for 7091 are T = 4200K, log(g) = 1.3, vt = 1.50 km s$^{-1}$}
\tablenotetext{c}{Starting parameters for 11622 are T = 4750K, log(g) = 2.3, vt = 1.45 km s$^{-1}$}
\label{atm_errors}
\end{deluxetable}

\indent We share stars 3835, 6810, 8734, and 10133 with JPF11; 6345, 6810, 7091, 8293, 8316, 8734, and 8799 with T05; and 5237 with PCRG10. Our spectroscopic temperatures compare well, with a typical difference of $\sim$ +50K from literature values. Our adopted GES gravities differ by approximately +0.2 dex compared to T05 and -0.2 dex compared to JPF11. Our log(g) for 5237 is 0.6 dex lower than found by PCRG10, but our resulting Fe I abundances only change by 0.02 dex. Our microturbulent velocities compare well with JPF11 and PCRG10 with differences $\le$0.1 km s$^{-1}$, but are typically 0.2 km s$^{-1}$ smaller than those of T05. Our Fe I abundances show rms deviations from the mean of about 0.1 dex, with a systematic increase compared to those found by T05.

\section{Abundance Determinations}

\subsection{Magnesium}

\indent We have measured three magnesium lines from the GES linelist at $\lambda$6318.72, 6319.24, and 6319.50$\mathrm{\AA}$. This wavelength range of 6317-6320$\mathrm{\AA}$ contains significant telluric contamination; we used our telluric-corrected spectra (see section 3) to measure these lines, assuming all contamination was removed. This region is also affected by a Ca autoionization feature, which changes the strength and shape of the Mg lines; thus, we have used spectral synthesis to obtain Mg abundances. We have used the linelist from Johnson et al. (2014), which includes Ca features, and set the stellar Ca abundance to match the shape of the autoionization feature/observed continuum.
\\
\indent Table 5 gives $\alpha$ element [X/Fe] ratios\footnote{Abundance ratios are expressed as logarithms relative to hydrogen, i.e., log $\epsilon$(X) = log$_{10}(N_X/N_H)$, [X/H] = log $\epsilon$(X)$_{\mathrm{star}}$ - log $\epsilon$(X)$_{\mathrm{\odot}}$, and [X/Fe] = [X/H] - [Fe/H]}, uncertainties in [X/Fe], and the number of lines used for each star. For Si, Ca, and Ti, the final [X/Fe] value is the average of [X/Fe] values for all lines, using each star's individual [Fe/H], and the given uncertainty is the standard deviation of those values. We omit errors for stellar abundances based on one line. For Mg, the stellar average is the result of fitting all three lines simultaneously with synthesis; the stellar error is the typical change in Mg abundance caused by a Ca abundance change of 0.20 dex - we judge 0.20 dex to be the smallest Ca change where the difference to the apparent continuum is clear given the S/N of our spectra. We have found a cluster average of [Mg/H] = 0.15 $\pm$ 0.07, and [Mg/Fe] = 0.11 $\pm$ 0.05.

\subsection{Silicon, Calcium, $\&$ Titanium}

\indent The GES linelist has no Si I lines with reliable log(gf) values in our wavelength range, so we have used the four lines from the Arcturus-based linelist instead. As discussed in section 4, there are no significant systematic differences between GES and Arcturus-based log(gf) values, so our Si measurements are comparable to other elemental abundances based on the GES linelist. Again, the equivalent widths measured from Si lines were not large, so only a handful of stars were analyzed based on fewer than four lines. For all elements measured, when running MOOG we trimmed any lines with abundances $> 2\sigma$ from the stellar average. We measured six Ca lines in our sample stars, but some lines ($\lambda$6102.72 and 6169.56) consistently gave equivalent widths $>$ 170 m$\mathrm{\AA}$ so our analyses are based on $\sim$4 lines per star. We find (weighted) cluster averages [Si/H] = 0.22 $\pm$ 0.07, [Si/Fe] = 0.17 $\pm$ 0.04,  [Ca/H] = 0.10 $\pm$ 0.10, [Ca/Fe] = 0.07 $\pm$ 0.08, [Ti/H] = 0.03 $\pm$ 0.12, [Ti/Fe] = -0.02 $\pm$ 0.10.

\begin{deluxetable*}{l c c c c c c c c c c c c c}
\tabletypesize{\small}
\tablewidth{0pt}
\tablecolumns{13}
\tablecaption{Abundances of $\alpha$-elements}
\tablehead{\colhead{ID\tablenotemark{a}} & \colhead{[Mg/Fe]} & \colhead{$\sigma_{\mathrm{Mg}}$} & \colhead{$n_{\mathrm{Mg}}$\tablenotemark{b}} & \colhead{[Si/Fe]} & \colhead{$\sigma_{\mathrm{Si}}$} & \colhead{$n_{\mathrm{Si}}$} & \colhead{[Ca/Fe]} & \colhead{$\sigma_{\mathrm{Ca}}$} & \colhead{$n_{\mathrm{Ca}}$} & \colhead{[Ti/Fe]} & \colhead{$\sigma_{\mathrm{Ti}}$} & \colhead{$n_{\mathrm{Ti}}$}}
\startdata
2075 & 0.11 & 0.15 & 3 & 0.16 & 0.11 & 4 & 0.18 & 0.14 & 5 & 0.00 & 0.09 & 7 \\
2427 & 0.01 & 0.15 & 3 & 0.28 & 0.11 & 4 & 0.11 & 0.04 & 3 & 0.03 & 0.12 & 8 \\
3798 & 0.12 & 0.15 & 3 & 0.19 & 0.15 & 4 & 0.08 & 0.13 & 3 & -0.01 & 0.10 & 8 \\
3835 & 0.06 & 0.15 & 3 & 0.13 & 0.08 & 4 & 0.03 & 0.09 & 4 & 0.08 & 0.10 & 9 \\
4593 & 0.03 & 0.15 & 3 & 0.19 & 0.10 & 4 & -0.09 & 0.13 & 4 & -0.22 & 0.05 & 7 \\
4751 & 0.12 & 0.15 & 3 & 0.16 & 0.01 & 2 & 0.09 & 0.08 & 5 & -0.05 & 0.13 & 9 \\
5237 & 0.11 & 0.15 & 3 & 0.18 & 0.09 & 4 & 0.08 & 0.15 & 6 & -0.02 & 0.13 & 8 \\
5594 & 0.13 & 0.15 & 3 & 0.16 & 0.13 & 4 & 0.00 & 0.07 & 5 & 0.04 & 0.13 & 9 \\
5837 & 0.11 & 0.15 & 3 & 0.22 & 0.07 & 4 & 0.04 & 0.09 & 3 & -0.06 & 0.11 & 8 \\
6810 & 0.08 & 0.15 & 3 & 0.14 & 0.05 & 4 & 0.19 & 0.03 & 2 & 0.23 & 0.13 & 6 \\
6863 & 0.12 & 0.15 & 3 & 0.17 & 0.04 & 4 & 0.10 & 0.13 & 4 & -0.08 & 0.16 & 9 \\
7091 & 0.10 & 0.15 & 3 & 0.19 & 0.06 & 3 & 0.11 & \nodata & 1 & 0.32 & 0.07 & 5 \\
7369 & 0.22 & 0.15 & 3 & 0.25 & 0.04 & 4 & 0.09 & 0.09 & 4 & -0.04 & 0.12 & 8 \\
7617 & 0.11 & 0.15 & 3 & 0.17 & 0.04 & 3 & 0.21 & 0.18 & 4 & 0.01 & 0.16 & 8 \\
7640 & 0.13 & 0.15 & 3 & 0.14 & 0.08 & 4 & -0.08 & 0.13 & 5 & -0.03 & 0.11 & 9 \\
7867 & 0.20 & 0.15 & 3 & 0.18 & 0.08 & 4 & 0.21 & 0.16 & 4 & 0.03 & 0.20 & 8 \\
8061 & 0.08 & 0.15 & 3 & 0.19 & 0.10 & 4 & 0.08 & 0.07 & 4 & -0.04 & 0.13 & 8 \\
8293 & 0.06 & 0.15 & 3 & 0.19 & 0.05 & 3 & -0.07 & 0.11 & 3 & -0.10 & 0.13 & 6 \\
8316 & 0.05 & 0.15 & 3 & 0.16 & 0.09 & 4 & 0.16 & 0.10 & 5 & 0.12 & 0.16 & 9 \\
8734 & 0.08 & 0.15 & 3 & 0.20 & 0.07 & 4 & 0.05 & 0.19 & 5 & -0.01 & 0.11 & 9 \\
9010 & 0.11 & 0.15 & 3 & 0.15 & 0.09 & 4 & 0.03 & 0.07 & 4 & 0.06 & 0.05 & 7 \\
10133 & 0.11 & 0.15 & 3 & 0.24 & 0.12 & 4 & 0.08 & 0.03 & 4 & -0.03 & 0.07 & 8 \\
10578 & 0.09 & 0.15 & 3 & 0.22 & 0.06 & 4 & 0.06 & 0.11 & 4 & 0.01 & 0.11 & 8 \\
10584 & 0.17 & 0.15 & 3 & 0.15 & 0.04 & 4 & -0.03 & 0.15 & 5 & -0.14 & 0.15 & 9 \\
10645 & 0.11 & 0.15 & 3 & 0.14 & 0.06 & 4 & 0.10 & 0.13 & 5 & -0.02 & 0.13 & 9 \\
10740 & 0.14 & 0.15 & 3 & 0.17 & 0.11 & 4 & 0.06 & 0.18 & 4 & -0.04 & 0.14 & 8 \\
10996 & 0.21 & 0.15 & 3 & 0.25 & 0.06 & 4 & 0.21 & 0.13 & 4 & 0.09 & 0.12 & 8 \\
11413 & 0.19 & 0.15 & 3 & 0.21 & 0.09 & 4 & 0.03 & 0.03 & 3 & -0.09 & 0.12 & 8 \\
11573 & 0.08 & 0.15 & 3 & 0.19 & 0.08 & 4 & 0.04 & 0.07 & 4 & -0.09 & 0.12 & 8 \\
11622 & 0.09 & 0.15 & 3 & 0.17 & 0.10 & 4 & 0.10 & 0.16 & 4 & -0.07 & 0.11 & 8 \\
12478 & 0.12 & 0.15 & 3 & 0.22 & 0.09 & 4 & 0.04 & 0.02 & 2 & -0.04 & 0.12 & 7 \\
12550 & 0.09 & 0.15 & 3 & 0.20 & 0.10 & 4 & 0.05 & 0.09 & 5 & -0.13 & 0.06 & 6 \\
\enddata
\tablenotetext{a}{ID numbers from Gim et al. (1998b), WEBDA.}
\tablenotetext{b}{All Mg lines were measured simultaneously via spectral synthesis.}
\label{alpha_abs}
\end{deluxetable*}

\subsection{Sodium, Nickel, $\&$ Zirconium}

\indent Literature measurements of [Na/Fe] ratios in open clusters vary from approximately solar to +0.50 dex or more (Smiljanic 2012). Some of this variation is likely due to the use of different lines and log(gf) values in measuring equivalent widths and calculating abundances, but non-LTE corrections also affect final abundances. Lind et al. (2011) find that the non-LTE corrections for the sodium $\lambda$6154/6160 doublet are small over the temperature and metallicity range of our sample stars ($\le$ 0.10 dex), and Mashonkina et al. (2000) similarly find that $\Delta$NLTE $\le$ 0.05 dex, so we do not apply any such correction. Table 6 shows [X/Fe], uncertainties, and number of lines used for all non-$\alpha$ elements. As with Table 5, given errors are the standard deviation of abundances from individual lines, and abundances based on a single line have no error listed. 
\\
\indent [Na/Fe] values from our equivalent width analysis show a slight trend with effective temperature. The slope of the best fit line is 8x10$^{-5}$ dex/K, or 0.07 dex within our sample temperature range. This 0.07 dex abundance range is smaller than our expected uncertainty due to atmospheric parameter errors (Table 4), so we present a cluster average [Na/Fe] = 0.25 $\pm$ 0.06. The possible implications of a Na trend with temperature/evolutionary state are discussed in section 7.5.
\\
\indent To check our equivalent width measurements, and to rule out blending as the cause of the trend, we also evaluated sodium abundances using MOOG spectral synthesis. We have used Autosynth, a code that automates MOOG synthesis measurements (M\'{e}sz\'{a}ros et al. 2014, in preparation), for uniformity and efficiency. Briefly, Autosynth reads in a spectrum, sets a continuum for each by fitting data points in a certain flux range, uses MOOG to generate synthetic spectra with a user-specified smoothing parameter for a range of input abundances, and selects the abundance that gives the lowest $\chi^2$ when comparing the synthetic spectrum with the data. For comparison purposes we have used a Na linelist that includes molecular features, courtesy of C. Sneden. The synthesis average [Na/Fe] = 0.16 $\pm$ 0.08, consistent with equivalent width-determined values within errors.
\\
\indent Our Autosynth results show a stronger trend of [Na/Fe] with T$_{\mathrm{eff}}$, changing by +0.20 dex from highest to lowest temperatures. Due to the larger temperature trend and dispersion in our synthesis measurements, we have included only [Na/Fe] results based on equivalent widths in Table 6.
\\
\indent The GES linelist contains 9 Ni lines in the wavelength range of our data; as with other elements, all lines with EQWs $>$ 170 m$\mathrm{\AA}$ or abundance values $>$ 2$\sigma$ from average were excluded. We find a cluster average of [Ni/Fe] = 0.04 $\pm$ 0.04. Finally, we measured equivalent widths of four Zr lines for our sample stars, and find a cluster average [Zr/Fe] = 0.08 $\pm$ 0.08 dex.

\subsection{Synthesis of Barium $\&$ Lanthanum}

Because barium and lanthanum are both subject to hyperfine broadening, we have used MOOG/Autosynth to measure both. We adopt the La $\lambda$6262 linelist from Lawler et al. (2001), because this feature has not yet been incorporated into the GES linelist. We did not measure C, N, or O in this study, so we adopt a cluster ratio [O/Fe] = 0.05 from JPF11, and [C, N/Fe] values of -0.21 and +0.09 based on measurements of NGC 7789 red clump stars in T05. We also use a $^{12}$C/$^{13}$C of 9 from T05. Our assumed CNO values affect only La abundances because only the $\lambda$6262 line is affected by CN features. For the Ba $\lambda$6141 feature, we have measured abundances based on both the GES linelist and the linelist of McWilliam (1998). The McWilliam linelist incorporates solar Ba isotope fractions from Anders and Grevesse (1989) into its log(gf) values; the GES linelist does not contain isotopic data for this feature.
\\
\indent Stellar [Ba/Fe] and [La/Fe] measurements and uncertainties are given in Table 6. We have estimated uncertainties as 3$\sigma$ of the residuals between the data and the synthetic spectrum. This is the typical size of uncertainties in abundances due to errors in atmospheric parameters (see Table 4), and what we estimate to be the smallest change in abundance from which we can reliably select a best fit given uncertainties in the continuum. We find a cluster abundance of [La/Fe] = -0.08 $\pm$ 0.05. Figure 6 shows a sample synthetic fit of the $\lambda$6262 La feature. The shape of the synthetic La line closely matches the observed spectrum, and the surrounding features, including CNO lines, are well-fit. This gives us confidence in our adopted CNO abundances.
\\
\indent Using the McWilliam (1998) linelist we find [Ba/Fe] = 0.48 $\pm$ 0.08, and using the GES linelist [Ba/Fe] = 0.47 $\pm$ 0.08.  Because the McWilliam Ba linelist is more comprehensive with regards to hyperfine splitting and isotopic data, we have chosen to report stellar abundances based on this linelist in Table 6. We have also used a high-resolution solar spectrum (Hinkle et al. 2000) to measure the solar Ba abundance with the McWilliam linelist - we find a high solar Ba abundance of 2.48 compared to a solar value of 2.13 reported by Anders \& Grevesse (1989). PCRG10 also use the McWilliam linelist for measurement of the Ba $\lambda$6141 feature, and they measure a solar abundance of 2.47; however, since the cause of this discrepancy is unclear, we follow PCRG10 and adopt the standard Anders \& Grevesse (1989) solar abundance for this paper. 
\\
\indent Figure 7 shows a synthetic fit of the $\lambda$6141 Ba feature in star 3835, with a solid line marking the best fit at Ba = +2.71, dashed lines showing typical Ba errors of 0.1 dex, and a dotted line showing a solar [Ba/Fe] ratio. The difference between our result and the Anders \& Grevesse (1989) solar Ba abundance is clearly too large to be attributed to random error. The variability of the solar Ba abundance from study to study suggests that Ba measurements are reliant on the technique and linelist used. That our Ba abundances agree so well with PCRG10, who derive theirs from equivalent width measurements, is encouraging, but we must interpret the results with some caution (see section 7.4). 

\begin{figure}
\epsscale{1.1}
\plotone{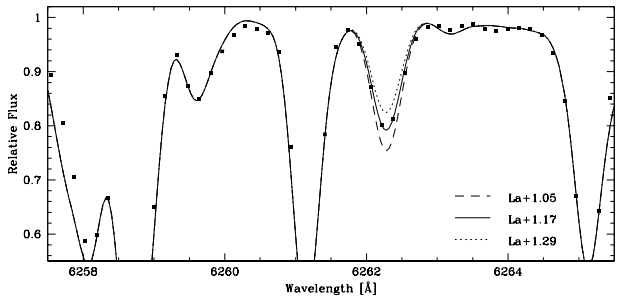}
\caption{\small{Comparison of the observed spectrum of star 12478 (dots) with the best spectral synthesis fit (solid line). The dashed and dotted lines indicate the uncertainties on the final [La/H] of $\pm$0.12 dex.}}
\label{La_synth}
\end{figure}

\begin{figure}
\epsscale{1.07}
\plotone{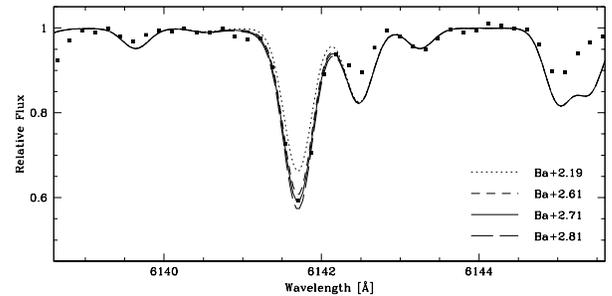}
\caption{\small{Comparison of the observed spectrum of star 3835 (dots) with the best spectral synthesis fit for Ba 6141 (solid line). The dashed lines indicate a typical stellar Ba uncertainty of $\pm$0.10 dex, and the dotted line shows a solar [Ba/Fe] ratio for this star.}}
\label{Ba_synth}
\end{figure}

\begin{deluxetable*}{l c c c c c c c c c c c c c c}
\tabletypesize{\small}
\tablewidth{0pt}
\tablecolumns{14}
\tablecaption{Other Abundances}
\tablehead{\colhead{ID\tablenotemark{a}} & \colhead{[Na/Fe]} & \colhead{$\sigma_{\mathrm{Na}}$} & \colhead{$n_{\mathrm{Na}}$} & \colhead{[Ni/Fe]} & \colhead{$\sigma_{\mathrm{Ni}}$} & \colhead{$n_{\mathrm{Ni}}$} & \colhead{[Zr/Fe]} & \colhead{$\sigma_{\mathrm{Zr}}$} & \colhead{$n_{\mathrm{Zr}}$} & \colhead{[Ba/Fe]} & \colhead{$\sigma_{\mathrm{Ba}}$} & \colhead{[La/Fe]} & \colhead{$\sigma_{\mathrm{La}}$}}
\startdata
2075 & 0.19 & 0.02 & 2 & 0.01 & 0.08 & 7 & 0.10 & 0.04 & 3 & 0.43 & 0.08 & -0.05 & 0.10 \\
2427 & 0.19 & \nodata & 1 & 0.15 & 0.22 & 7 & 0.06 & 0.01 & 2 & 0.44 & 0.09 & -0.06 & 0.13 \\
3798 & 0.16 & 0.02 & 2 & 0.05 & 0.11 & 7 & 0.00 & 0.02 & 4 & 0.56 & 0.09 & 0.01 & 0.11 \\
3835 & 0.17 & 0.10 & 2 & 0.11 & 0.10 & 7 & 0.16 & 0.05 & 3 & 0.52 & 0.08 & -0.01 & 0.10 \\
4593 & 0.18 & 0.06 & 2 & 0.06 & 0.07 & 7 & 0.10 & 0.15 & 4 & 0.42 & 0.08 & -0.14 & 0.11 \\
4751 & 0.24 & 0.03 & 2 & 0.07 & 0.13 & 7 & 0.19 & 0.09 & 3 & 0.42 & 0.09 & -0.14 & 0.10 \\
5237 & 0.26 & 0.01 & 2 & 0.00 & 0.13 & 8 & 0.13 & 0.17 & 3 & 0.41 & 0.08 & -0.16 & 0.09 \\
5594 & 0.19 & 0.02 & 2 & 0.13 & 0.11 & 7 & 0.21 & 0.05 & 3 & 0.46 & 0.07 & -0.07 & 0.08 \\
5837 & 0.28 & 0.02 & 2 & 0.03 & 0.10 & 8 & 0.05 & 0.09 & 3 & 0.53 & 0.10 & -0.09 & 0.11 \\
6810 & 0.19 & 0.06 & 2 & 0.07 & 0.11 & 5 & 0.24 & 0.06 & 4 & 0.59 & 0.14 & -0.12 & 0.14 \\
6863 & 0.17 & 0.07 & 2 & 0.01 & 0.11 & 7 & 0.07 & 0.09 & 3 & 0.46 & 0.08 & -0.08 & 0.10 \\
7091 & 0.34 & 0.06 & 2 & 0.05 & 0.18 & 5 & 0.28 & 0.10 & 4 & 0.66 & 0.15 & 0.00 & 0.16 \\
7369 & 0.30 & 0.03 & 2 & 0.09 & 0.11 & 6 & 0.09 & 0.12 & 4 & 0.60 & 0.10 & -0.10 & 0.11 \\
7617 & 0.31 & 0.06 & 2 & 0.03 & 0.10 & 8 & 0.06 & 0.03 & 3 & 0.55 & 0.10 & -0.02 & 0.10 \\
7640 & 0.20 & 0.04 & 2 & 0.01 & 0.07 & 7 & 0.17 & 0.06 & 3 & 0.39 & 0.08 & -0.04 & 0.09 \\
7867 & 0.23 & 0.09 & 2 & 0.07 & 0.10 & 7 & 0.18 & 0.06 & 4 & 0.49 & 0.10 & -0.07 & 0.11 \\
8061 & 0.20 & 0.02 & 2 & -0.01 & 0.08 & 7 & 0.10 & 0.01 & 3 & 0.39 & 0.09 & -0.18 & 0.11 \\
8293 & 0.28 & 0.02 & 2 & 0.08 & 0.12 & 6 & 0.00 & 0.08 & 4 & 0.55 & 0.11 & -0.15 & 0.12 \\
8316 & 0.37 & 0.01 & 2 & 0.03 & 0.09 & 7 & 0.25 & 0.05 & 3 & 0.35 & 0.09 & -0.12 & 0.10 \\
8734 & 0.23 & 0.02 & 2 & 0.00 & 0.10 & 8 & 0.08 & 0.08 & 3 & 0.52 & 0.09 & -0.13 & 0.10 \\
9010 & 0.16 & 0.03 & 2 & 0.02 & 0.12 & 7 & 0.20 & 0.07 & 3 & 0.50 & 0.08 & -0.01 & 0.09 \\
10133 & 0.27 & 0.01 & 2 & 0.04 & 0.11 & 7 & 0.08 & 0.07 & 3 & 0.54 & 0.08 & 0.01 & 0.09 \\
10578 & 0.26 & 0.04 & 2 & 0.04 & 0.09 & 6 & 0.11 & 0.06 & 3 & 0.46 & 0.08 & -0.06 & 0.10 \\
10584 & 0.17 & 0.06 & 2 & 0.10 & 0.13 & 7 & 0.08 & 0.15 & 4 & 0.56 & 0.09 & -0.05 & 0.11 \\
10645 & 0.16 & 0.03 & 2 & -0.04 & 0.10 & 8 & 0.01 & 0.07 & 3 & 0.35 & 0.08 & -0.14 & 0.10 \\
10740 & 0.18 & 0.07 & 2 & 0.06 & 0.05 & 6 & 0.09 & 0.11 & 4 & 0.48 & 0.09 & -0.14 & 0.11 \\
10996 & 0.33 & 0.04 & 2 & 0.05 & 0.10 & 7 & 0.22 & 0.10 & 4 & 0.57 & 0.11 & -0.03 & 0.10 \\
11413 & 0.27 & 0.07 & 2 & 0.11 & 0.09 & 7 & -0.02 & 0.10 & 4 & 0.58 & 0.10 & -0.02 & 0.11 \\
11573 & 0.23 & 0.09 & 2 & -0.04 & 0.10 & 8 & 0.01 & 0.12 & 4 & 0.52 & 0.08 & -0.14 & 0.10 \\
11622 & 0.26 & 0.01 & 2 & 0.03 & 0.12 & 8 & 0.05 & 0.06 & 3 & 0.46 & 0.09 & -0.07 & 0.10 \\
12478 & 0.30 & 0.08 & 2 & 0.08 & 0.14 & 6 & 0.06 & 0.09 & 4 & 0.59 & 0.11 & -0.08 & 0.12 \\
12550 & 0.17 & 0.01 & 2 & 0.04 & 0.12 & 7 & -0.03 & 0.13 & 4 & 0.48 & 0.08 & -0.09 & 0.11 \\
\enddata
\tablenotetext{a}{ID numbers from Gim et al. (1998b), WEBDA.}
\label{other_abs}
\end{deluxetable*}

\section{Discussion and Comparison with Literature Results}
\subsection{Abundance Scales}

\indent Table 7 gives weighted average values of [X/Fe], uncertainties, literature cluster averages placed on our abundance scale, and solar abundances used for this study. Column 1 averages are weighted by stellar errors, and the error on the cluster abundance is the standard deviation of all stellar measurements. Columns 2 and 4 are the JPF11 and PCRG10 abundances placed on our abundance scale, i.e. adjusted so that they are relative to our adopted solar abundances (column 6). This was not done with T05 cluster averages because their solar abundance measurements are not available. Our study, JPF11, T05, and PCRG10 all place the Fe abundance of NGC 7789 at around solar, and the cluster averages for Fe agree within the errors. Table 8 shows abundance differences between our study and the literature for individual stars in common; values are on our abundance scale for the 5 stars our sample shares with JPF11 and PCRG10. All values for the literature are differential for comparison purposes $(\Delta$[X/Fe] = [X/Fe]$_{\mathrm{lit}}$ - [X/Fe]$_{\mathrm{here}})$. Figure 8 displays the information in Table 8 as a plot of abundances for all overlapping sample stars; abundances from this study are filled black circles, from JPF11 are open green squares, from PCRG10 are open red circles, and from T05 are open blue triangles. Measurements of the same star are plotted next to each other.

\begin{deluxetable*}{l c c c c c c}
\tablewidth{0pt}
\tablecolumns{6}
\tablecaption{Weighted Cluster Averages Compared to Literature Values}
\tablehead{\colhead{El.} & \colhead{Cluster Avg.} & \colhead{JPF11} & \colhead{T05} & \colhead{PCRG10} & \colhead{log($\epsilon$)}}
\startdata
[Fe/H] & 0.03 $\pm$ 0.07 & 0.02 $\pm$ 0.04 & -0.04 $\pm$ 0.05 & 0.02 $\pm$ 0.07 & 7.52 \\
{[}Mg/Fe] & 0.11 $\pm$ 0.05 & 0.14 $\pm$ 0.05 & 0.18 $\pm$ 0.07 & 0.24 $\pm$ 0.07 & 7.58 \\
{[}Si/Fe] & 0.17 $\pm$ 0.04 & 0.25 $\pm$ 0.05 & 0.14 $\pm$ 0.05 & 0.01 $\pm$ 0.02 & 7.55 \\
{[}Ca/Fe] & 0.07 $\pm$ 0.08 & 0.01 $\pm$ 0.05 & 0.14 $\pm$ 0.07 & -0.16 $\pm$ 0.09 & 6.36 \\
{[}Ti/Fe] & -0.02 $\pm$ 0.10 & -0.05 $\pm$ 0.04 & -0.03 $\pm$ 0.07 & 0.02 $\pm$ 0.09 & 4.99 \\
{[}Na/Fe] & 0.25 $\pm$ 0.06 & 0.09 $\pm$ 0.05 & 0.28 $\pm$ 0.07 & -0.03 $\pm$ 0.13 & 6.33 \\
{[}Ni/Fe] & 0.04 $\pm$ 0.04 & 0.00 $\pm$ 0.05 & -0.02 $\pm$ 0.05 & 0.01 $\pm$ 0.01 & 6.25 \\
{[}Zr/Fe] & 0.08 $\pm$ 0.08 & 0.18 $\pm$ 0.05 & -0.02 $\pm$ 0.13 & \nodata & 2.60 \\
{[}Ba/Fe] & 0.48 $\pm$ 0.08 & \nodata & \nodata & 0.47 $\pm$ 0.05 & 2.13 \\
{[}La/Fe] & -0.08 $\pm$ 0.05 & \nodata & \nodata & 0.08 $\pm$ 0.05 & 1.22 \\
No. stars & 32 & 28 & 9 & 3 & \nodata \\
\enddata
\label{cluster_avg}
\end{deluxetable*}

\indent The differences in stellar Fe abundance between studies are sometimes substantial; our values are systematically higher than T05 by about 0.1 dex for the five shared stars, in keeping with the 0.07 dex difference in the cluster average. Differences with JPF11 are also significant ($\sim$0.1 dex) and do not appear to be systematic; however, it is worth noting that our overlap with T05 contains 5 out of 9 stars, whereas our overlap with JPF11 is only 4 out of 32 stars. The differences in shared stars with JPF11, compared to overall cluster averages, may be more susceptible to small number statistics. It is also worth noting that JPF11 do not refine atmospheric parameters spectroscopically due to small numbers of Fe lines; this may cause some variation in Fe abundances. Presumably, any such effects are not present in [X/Fe] ratios.

\begin{figure*}
\epsscale{1.0}
\plotone{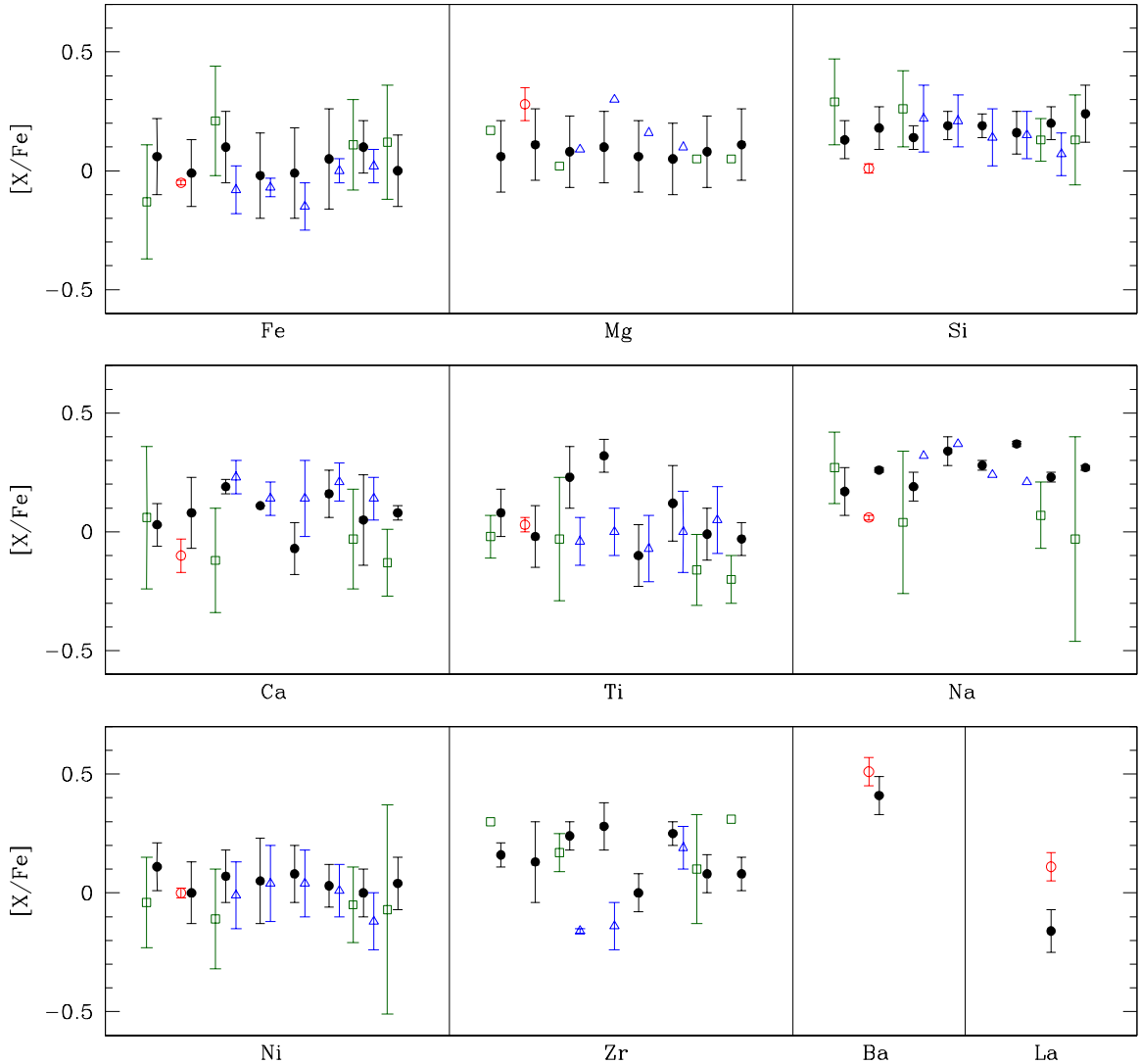}
\caption{\small{Comparison of literature abundances for all overlapping sample stars. Black filled circles are abundances from this study, green open squares are from JPF11, red open circles are from PCRG10, and blue open triangles are from T05. Measurements of the same star are placed next to each other. Stars that are not assigned errors in the literature are shown without error bars.}}
\label{lit_comp}
\end{figure*}

\begin{deluxetable*}{l c c c c c c c c c c c c}
\tablewidth{0pt}
\tablecolumns{12}
\tablecaption{Stellar Abundances Compared to Literature Values}
\tablehead{\colhead{Source} & \colhead{ID} & \colhead{$\Delta$[Fe/H]} & \colhead{$\Delta$[Mg/Fe]} & \colhead{$\Delta$[Si/Fe]} & \colhead{$\Delta$[Ca/Fe]} & \colhead{$\Delta$[Ti/Fe]} & \colhead{$\Delta$[Na/Fe]} & \colhead{$\Delta$[Ni/Fe]} & \colhead{$\Delta$[Zr/Fe]} & \colhead{$\Delta$[Ba/Fe]} & \colhead{$\Delta$[La/Fe]}}
\startdata
JPF11 & 3835 & -0.19 & +0.11 & +0.16 & +0.03 & -0.10 & +0.10 & -0.15 & +0.14 & \nodata & \nodata \\
 & 6810 & +0.11 & -0.09 & +0.12 & -0.31 & -0.26 & -0.15 & -0.18 & -0.07 & \nodata & \nodata \\
 & 8734 & +0.01 & -0.03 & -0.07 & -0.08 & -0.15 & -0.16 & -0.05 & +0.02 & \nodata & \nodata \\
 & 10133 & +0.12 & -0.06 & -0.11 & -0.21 & -0.17 & -0.30 & -0.11 & +0.23 & \nodata & \nodata \\
\hline
T05 & 6810 & -0.18 & +0.01 & +0.08 & +0.04 & -0.27 & +0.13 & -0.08 & -0.40 & \nodata & \nodata \\
 & 7091 & -0.05 & +0.20 & +0.02 & +0.03 & -0.32 & +0.03 & -0.01 & -0.42 & \nodata & \nodata \\
 & 8293 & -0.14 & +0.10 & -0.05 & +0.21 & +0.03 & -0.04 & -0.04 & \nodata & \nodata & \nodata \\
 & 8316 & -0.05 & +0.05 & -0.01 & +0.05 & -0.12 & -0.16 & -0.02 & -0.06 & \nodata & \nodata \\
 & 8734 & -0.08 & \nodata & -0.13 & +0.09 & +0.06 & \nodata & -0.12 & \nodata & \nodata & \nodata \\
\hline
PCRG10 & 5237 & -0.02 & +0.17 & -0.13 & -0.16 & +0.05 & -0.20 & 0.00 & \nodata & +0.10 & +0.27 \\
\hline
Source & ID& [Fe/H] & [Mg/Fe] & [Si/Fe] & [Ca/Fe] & [Ti/Fe] & [Na/Fe] & [Ni/Fe] & [Zr/Fe] & [Ba/Fe] & [La/Fe] \\
\hline
This Study & 3835 & 0.06 & 0.06 & 0.13 & 0.03 & 0.08 & 0.17 & 0.11 & 0.16 & 0.52 & -0.01 \\
 & 5237 & -0.01 & 0.11 & 0.18 & 0.08 & -0.02 & 0.26 & 0.00 & 0.13 & 0.41 & -0.16 \\
 & 6810 & 0.10 & 0.08 & 0.14 & 0.19 & 0.23 & 0.38 & 0.07 & 0.24 & 0.59 & -0.12 \\
 & 7091 & -0.02 & 0.10 & 0.19 & 0.11 & 0.32 & 0.35 & 0.05 & 0.28 & 0.66 & 0.00 \\
 & 8293 & -0.01 & 0.06 & 0.19 & -0.07 & -0.10 & 0.28 & 0.08 & 0.00 & 0.55 & -0.15 \\
 & 8316 & 0.05 & 0.05 & 0.16 & 0.16 & 0.12 & 0.37 & 0.03 & 0.25 & 0.35 & -0.12 \\
 & 8734 & 0.10 & 0.08 & 0.20 & 0.05 & -0.01 & 0.23 & 0.00 & 0.08 & 0.52 & -0.13 \\
 & 10133 & 0.00 & 0.11 & 0.24 & 0.08 & -0.03 & 0.27 & 0.04 & 0.08 & 0.54 & 0.01 \\
\enddata
\label{star_lit}
\end{deluxetable*}

\subsection{$\alpha$ Elements}

\indent Our average [Mg/Fe] abundances agree with JPF11, and are systematically low compared to T05. This is possibly due to differences in linelists - T05 use one Mg line at $\sim$8700$\mathrm{\AA}$ and JPF11 use the $\lambda$6319.24 line with a 0.08 dex higher log(gf) value. It is also worth noting that JPF11 account for the Ca autoionization feature by altering the continuum in that region.
\\
\indent In order to gauge the effect of the treatment of the Ca feature on our Mg abundances, we have also measured the three Mg lines in the GES linelist using equivalent widths but not compensating for the effect of Ca on the local continuum, similar to the way the EPINARBO group currently measures stellar Mg abundances. Using the stellar average of abundances from equivalent width measurements of all three lines and and the standard deviation for stellar errors, we find a (weighted) cluster average of [Mg/H] = 0.31 $\pm$ 0.08, and [Mg/Fe] = 0.29 $\pm$ 0.05. First measurements of open clusters from GES data are sometimes yielding [Mg/Fe] ratios up to 0.2 dex higher than those for other $\alpha$ elements (Magrini et al. 2014); an unfitted Ca feature could be causing this apparent Mg enhancement.
\\
\\
\indent Though individual stellar differences in [Ti/Fe] between studies are at times significant ($\sim$0.15 dex), they generally fall within the errors on each set of measurements. Cluster [Ti/Fe] averages are all solar or slightly sub-solar and agree well. [Si/Fe] and [Ca/Fe] measurements display systematic offsets between studies; our cluster averages fall between those of T05 and JPF11 with PCRG10 giving the lowest values for both abundances. For Ca, these differences could be due to log(gf) values; PCRG10 note that there are large differences (up to 0.2 dex) between Ca log(gf) values in the literature.
\\
\indent For Si values, the discrepancies may be caused by the use of lines in different wavelength regimes. We use the same lines as JPF11 and very similar log(gf) values and find similar abundances. We share a couple of lines with T05 and have comparable results, but PCRG10 use a completely different set of Si lines. We also note that PCRG10 use DAOSPEC to measure equivalent widths, and DAOSPEC employs a different continuum-fitting method than traditionally used in stellar spectroscopy - it finds the \textit{effective continuum} by compensating for a blanketing of small lines in spectra. This generally leads to a lower equivalent width measurement than other methods, particularly in metal-rich giants (Stetson $\&$ Pancino 2008).

\subsection{Sodium $\&$ Nickel}

\indent Stellar [Ni/Fe] abundances agree well, typically within 0.1 dex, although our measurements are systematically slightly higher than comparison studies. Cluster averages are all about solar, with differences $<$ 0.10 dex, and all consistent within errors.
\\
\indent [Na/Fe] values show discrepancies of up to 0.3 dex between both stellar measurements and cluster averages; our values are most similar to those of T05. The cause of these differences is unclear. We and JPF11 and PCRG10 use the same doublet with similar log(gf) values, although PCRG10 also use the $\lambda$5688 Na feature which has been found to give lower abundances than $\lambda$6154 and 6160 lines (Friel et al. 2003, Bragaglia et al. 2001). Equivalent widths for these lines are roughly comparable with no apparent systematic differences. T05 use one line, $\lambda$6154, for equivalent width measurements with similar abundance results. Furthermore, our errors in abundances due to uncertainties in atmospheric parameters (Table 4) indicate that Fe and Na respond similarly to changes over the size of the differences in parameters between this study and JPF11. However, [Na/Fe] is often found to be super-solar in open clusters, and our value is not unusual for clusters of NGC 7789's age and galactocentric distance  (e.g., Friel et al. 2010, Smiljanic 2012). 

\subsection{Neutron-capture Elements}

\indent JPF11 and T05 give Zr abundances for NGC 7789; our cluster result overlaps with both within the combined errors, but there are still significant differences in star-to-star measurements as high as 0.4 dex. Differences between our and JPF11's results may be due to line selection - the GES linelist includes 4 Zr lines, and JPF11 measure 2 of these features. T05 use three of our four Zr lines with similar equivalent widths, but lower values may be due to atmospheric parameter differences; errors in Table 4 suggest that Zr is quite sensitive to T$_{\mathrm{eff}}$. 
\\
\indent The only previous measurements of Ba and La for NGC 7789 are from PCRG10. Our [Ba/Fe] ratios for 5237 and the cluster as a whole agree well. Our [La/Fe] values are discrepant; this could be due to the fact that we measured different lines, or because we have used different methods. Our La abundances are based on a single feature, $\lambda$6262, and we used synthesis with a line list incorporating hyperfine structure. PCRG10 measure three La features - $\lambda$5290, 6390, and 6774 - as equivalent widths. Because some structure is visible in the $\lambda$5290 and 6262 features of the Hinkle et al. (2000) high-resolution Arcturus spectrum, we expect metal-rich open cluster stars to also display this structure.
\\
\indent Maiorca et al. (2011) measured La and Zr abundances for six open clusters with ages from $\sim$0.5 - 4 Gyr. The cluster in their sample closest in age to NGC 7789 is the 1.7 Gyr old OC IC 4651; they find a [Zr/Fe] = 0.08 $\pm$ 0.03 and [La/Fe] = 0.07 $\pm$ 0.03 for IC 4651. These results come from a limited sample (5 stars in IC 4651), but the [Zr/Fe] ratio matches what we find for NGC 7789 and fits the distribution with age from Jacobson \& Friel (2013).
\\
\indent [Ba/Fe] has been found to increase dramatically with decreasing cluster age (D'Orazi et al. 2009). Our measurement for [Ba/Fe] is consistent with the trend in [Ba/Fe] seen in D'Orazi et al. based on measurements of giants in open clusters at ages of 1-2 Gyr. These highly enhanced Ba abundances cannot be explained with current models of stellar evolution and mixing; models previously predicted a plateau or even a decrease of [Ba/Fe] as stars approach solar metallicity (Raiteri et al. 1999, Travaglio et al. 1999, Busso et al. 2001). 
\\
\indent These recent data require enhanced s-process production in low-mass (1-1.5 M$_{\odot}$) AGB stars with highly effective mixing (D'Orazi et al. 2009, Maiorca et al. 2012). Because low-mass AGB stars have only recently begun to pollute the ISM, only the youngest clusters would have formed from super-enriched gas. However, [La/Fe] shows only a slight trend with cluster age, and [Zr/Fe] shows some increases with decreasing age but at only half the rate of Ba (Jacobson \& Friel 2013). It is still unclear why abundance trends should vary for these three neutron-capture elements.

\subsection{Elemental Trends with Evolutionary State}

\indent Because we have abundance determinations for 32 stars, a larger sample than most OC studies, we are able to look for trends in abundances with evolutionary state. It has been suggested that Na produced by the $^{22}$Ne ($p, \gamma)^{23}$Na reaction in the interiors of red giants is then brought to the surface of the star during the first dredge-up (e.g. Sweigart \& Mengel 1979, Denisenkov \& Denisenkova 1990, Denissenkov \& VandenBerg 2003, Smiljanic 2012). Depending on the mixing parameters used, one might expect that cooler stars with deeper convection zones should show an increase in Na relative to hotter cluster stars. 
\\
\indent Figure 9 shows average [Na/Fe] abundances compared to [Ba/Fe], [Ni/Fe], [Ca/Fe], [La/Fe] and [Fe/H] for stars split into three V magnitude bins: V $>$ 12.7 (clump 
\\
\\
stars, 19), 11.8 $<$ V $<$ 12.7 (lower RGB, 9), and V $<$ 11.8 (upper RGB, 4). Horizontal and vertical error bars show the standard deviations in V and [X/Fe], respectively. [Ca/Fe], [Ni/Fe], and [La/Fe] show no trend with V magnitude or luminosity, as expected. [Na/Fe] possibly shows a weak trend with V, but the error bars in each of the bins overlap, and the dispersion on the cluster average is consistent with the errors due to atmospheric parameters (see Table 4 and Table 7). In fact, [Ba/Fe] shows a stronger trend than Na, although the cluster dispersion is again within estimated uncertainties. The [Ba/Fe] trend is unlikely to be real; neutron-capture abundances are not thought to be strongly affected by the first dredge-up phase.

\begin{figure}
\epsscale{1.0}
\plotone{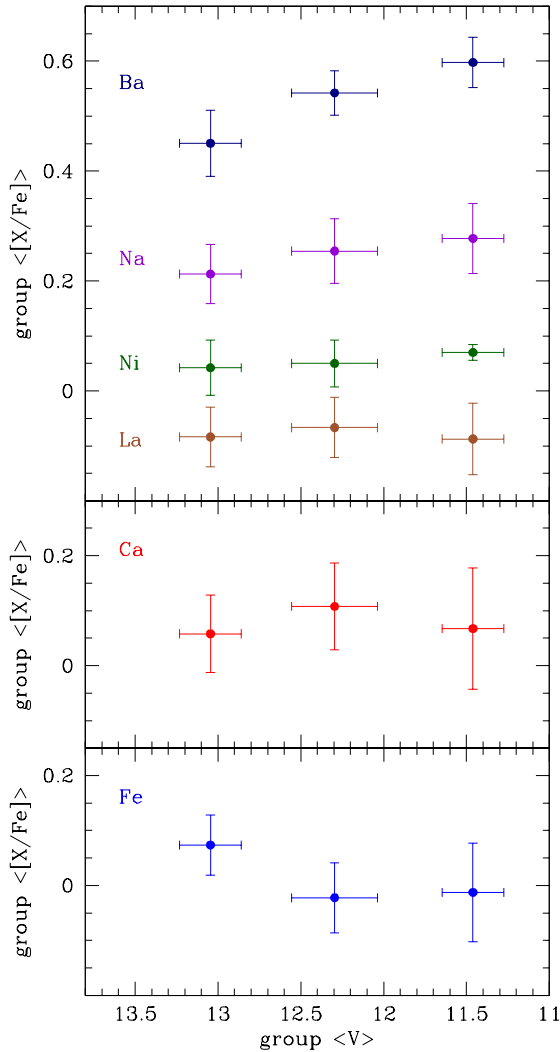}
\caption{\small{Average [X/Fe] values for stars in three V magnitude bins, V $>$ 12.7, 11.8 $<$ V $<$ 12.7, and V $<$ 11.8. Error bars show the standard deviation of [X/Fe] and V within the group.}}
\label{Na_sig}
\end{figure}

\section{Summary and Conclusions}

We have presented radial velocities and abundance measurements of ten elements for 32 giant stars in open cluster NGC 7789; as far as we know, this is the largest existing sample size for abundance analyses of this cluster. The Gaia-ESO Survey (GES), a large spectroscopic survey of stars in multiple components of the Milky Way that will greatly enhance future studies of Galactic chemical enrichment, has recently begun placing standard stars on its abundance scale for comparison with literature measurements. With this in mind, we have spectroscopically refined stellar atmospheric parameters and calculated stellar Fe abundances using two different linelists - a linelist developed as the standard for the GES, and a linelist using Arcturus-based log(gf) values from Jacobson et al. (2011). 
\\
\indent We find that while temperatures and microturbulent velocities determined using the two linelists compare well, gravities based on the GES linelist are systematically lower by $\sim$0.1 dex than gravities from the Arcturus-based linelist. This difference in gravities drives a systematic difference in Fe abundances; using the GES linelist, we find a cluster [Fe/H] of 0.03 $\pm$ 0.07 dex, whereas the Arcturus-based linelist gives a cluster [Fe/H] = 0.11 $\pm$ 0.09 dex. 
\\
\indent We find most cluster abundances to be $\sim$solar (within 0.1 dex), except for [Na/Fe] = 0.25 $\pm$ 0.06, and [Ba/Fe] = 0.48 $\pm$ 0.08. Differences between abundances using the GES linelist and those found in the literature were expected in some cases, particularly Mg due to the treatment of the Ca autoionization feature in the same region, and measurement of different lines and differences in log(gf) values (although these generally compare well; see Figure 4). The fact that Zr and La measurements are so different from Ba is interesting and agrees with other recent studies of neutron-capture elements in open clusters that challenge current models of neutron-capture production in AGB stars (D'Orazi et al. 2009, Jacobson \& Friel 2013). Also, the semi-automated synthesis program Autosynth seems to yield consistent results (cluster $\sigma <$ 0.10 dex), and provides a method for efficiently measuring larger numbers of clusters and/or more lines. Further investigation of neutron-capture elements in a large and homogenous open cluster sample may resolve the conflict between theory and observation.

\acknowledgments
C. I. J. gratefully acknowledges support from the Clay Fellowship, administered by the Smithsonian Astrophysical Observatory. C.A.P. acknowledges the generosity of the Kirkwood Research Fund at Indiana University. We would also like to thank WIYN telescope operator George Will for all of his help. This research has made use of the WEBDA database, operated at the Institute for Astronomy of the University of Vienna, and data products from the 2MASS, which is a joint project of the University of Massachusetts and the Infrared Processing and Analysis Center/California Institute of Technology funded by NASA and the NSF.

{\it Facilities:} \facility{WIYN (Hydra)}.

\end{document}